\documentclass[trackchanges,twocolumn]{aastex701}

\usepackage{amsmath}

\begin{document}

\title{Influence of mass-transfer stability on the formation of post-common-envelope binaries}

\author[orcid=0009-0001-2504-7054,gname='Yanxu', sname='Shi']{Yanxu Shi}
\affiliation{Yunnan Observatories, Chinese Academy of Sciences (CAS), Kunming 650216, People's Republic of China}
\affiliation{International Centre of Supernovae, Yunnan Key Laboratory, Kunming 650216, People's Republic of China}
\affiliation{University of Chinese Academy of Sciences, Beijing 100049, People's Republic of China}
\email{2675223153@qq.com}  

\author[orcid=0000-0002-6398-0195,gname='Hongwei', sname='Ge']{Hongwei Ge}
\affiliation{Yunnan Observatories, Chinese Academy of Sciences (CAS), Kunming 650216, People's Republic of China}
\affiliation{International Centre of Supernovae, Yunnan Key Laboratory, Kunming 650216, People's Republic of China}
\affiliation{University of Chinese Academy of Sciences, Beijing 100049, People's Republic of China}
\email{gehw@ynao.ac.cn}

\author[orcid=0000-0002-1421-4427,gname='Zhenwei', sname='Li']{Zhenwei Li}
\affiliation{Yunnan Observatories, Chinese Academy of Sciences (CAS), Kunming 650216, People's Republic of China}
\affiliation{International Centre of Supernovae, Yunnan Key Laboratory, Kunming 650216, People's Republic of China}
\affiliation{University of Chinese Academy of Sciences, Beijing 100049, People's Republic of China}
\email{.}

\author[orcid=0000-0003-1535-0866,gname='Diogo', sname='Belloni']{Diogo Belloni}
\affiliation{Yunnan Observatories, Chinese Academy of Sciences (CAS), Kunming 650216, People's Republic of China}
\affiliation{International Centre of Supernovae, Yunnan Key Laboratory, Kunming 650216, People's Republic of China}
\email{.}

\author[orcid=0009-0001-6159-2236,gname='Rizhong', sname='Zheng']{Rizhong Zheng}
\affiliation{Yunnan Observatories, Chinese Academy of Sciences (CAS), Kunming 650216, People's Republic of China}
\affiliation{International Centre of Supernovae, Yunnan Key Laboratory, Kunming 650216, People's Republic of China}
\affiliation{University of Chinese Academy of Sciences, Beijing 100049, People's Republic of China}
\email{.}

\author[orcid=0000-0003-4265-7783,gname='Dengkai', sname='Jiang']{Dengkai Jiang}
\affiliation{Yunnan Observatories, Chinese Academy of Sciences (CAS), Kunming 650216, People's Republic of China}
\affiliation{International Centre of Supernovae, Yunnan Key Laboratory, Kunming 650216, People's Republic of China}
\affiliation{University of Chinese Academy of Sciences, Beijing 100049, People's Republic of China}
\email{dengkai@ynao.ac.cn}

\author[orcid=0009-0006-9211-2860,gname='Hailiang', sname='Chen']{Hailiang Chen}
\affiliation{Yunnan Observatories, Chinese Academy of Sciences (CAS), Kunming 650216, People's Republic of China}
\affiliation{International Centre of Supernovae, Yunnan Key Laboratory, Kunming 650216, People's Republic of China}
\affiliation{University of Chinese Academy of Sciences, Beijing 100049, People's Republic of China}
\email{chenhl@ynao.ac.cn}

\author[orcid=0009-0004-2396-7174,gname='A.', sname='Santos-Garcia']{A. Santos-Garcia}
\affiliation{Departament de Física, Universitat Politècnica de Catalunya, c/ Esteve Terrades 5, 08860, Castelldefels, Barcelona, Spain}
\email{.}

\author[orcid=0000-0001-5777-5251,gname='S.', sname='Torres']{S. Torres}
\affiliation{Departament de Física, Universitat Politècnica de Catalunya, c/ Esteve Terrades 5, 08860, Castelldefels, Barcelona, Spain}
\affiliation{Institut d’Estudis Espacials de Catalunya, Esteve Terrades, 1, Edifici RDIT, Campus PMT-UPC, 08860, Castelldefels, Barcelona, Spain}
\email{.}

\author[orcid=0000-0002-6153-7173,gname='A.', sname='Rebassa-Mansergas']{A. Rebassa-Mansergas}
\affiliation{Departament de Física, Universitat Politècnica de Catalunya, c/ Esteve Terrades 5, 08860, Castelldefels, Barcelona, Spain}
\affiliation{Institut d’Estudis Espacials de Catalunya, Esteve Terrades, 1, Edifici RDIT, Campus PMT-UPC, 08860, Castelldefels, Barcelona, Spain}
\email{.}

\author[orcid=0000-0001-5284-8001,gname='Xuefei', sname='Chen']{Xuefei Chen}
\affiliation{Yunnan Observatories, Chinese Academy of Sciences (CAS), Kunming 650216, People's Republic of China}
\affiliation{International Centre of Supernovae, Yunnan Key Laboratory, Kunming 650216, People's Republic of China}
\affiliation{University of Chinese Academy of Sciences, Beijing 100049, People's Republic of China}
\email{.}

\author[orcid=0000-0001-9204-7778,gname='Zhanwen', sname='Han']{Zhanwen Han}
\affiliation{Yunnan Observatories, Chinese Academy of Sciences (CAS), Kunming 650216, People's Republic of China}
\affiliation{International Centre of Supernovae, Yunnan Key Laboratory, Kunming 650216, People's Republic of China}
\affiliation{University of Chinese Academy of Sciences, Beijing 100049, People's Republic of China}
\email{.}

\correspondingauthor{gehw@ynao.ac.cn}



\begin{abstract}
Post-common-envelope binaries are the natural laboratories for constraining the physics of common-envelope evolution, which is one of the most uncertain phases in binary stellar evolution. Traditional binary population synthesis models, adopting mass-transfer stability criteria based on polytropic stellar models, systematically overpredict the number of post-common-envelope binaries with solar-type main-sequence companions. In this work, we present an updated binary population synthesis model using the rapid binary evolution code \textit{binary star evolution}, incorporating a physically motivated mass-transfer stability criterion and a self-consistent envelope binding-energy prescription. We compile a comprehensive sample of classic white dwarf + main-sequence post-common-envelope binaries with well-measured parameters, hosting both M dwarf and A/F/G/K stars. We find that the enhanced mass-transfer stability is an additional mechanism responsible for the observed dearth of post-common-envelope binaries with solar-type main-sequence companions; neither magnetic braking nor selection effects alone can fully account for this deficit, and a combination of all three processes is most likely required. Models with inefficient common-envelope evolution ($\alpha_{\rm CE}=0.25$) provide the best overall match to the observed population. These results highlight the critical role of mass-transfer stability in shaping the observed post-common-envelope binary population and provide new constraints on common-envelope evolution.
\end{abstract}


\keywords{\uat{Common envelope evolution}{2154} --- \uat{Common envelope binary stars}{2156} --- \uat{White dwarf stars}{1799} --- \uat{Binary stars}{154} --- \uat{Stellar evolution}{1599}}


\section{Introduction}

A large fraction of all stars in the Galaxy reside in binary systems \citep{2013ARA&A..51..269D,2017ApJS..230...15M,2024PrPNP.13404083C}, which are the progenitors of the majority of exotic stellar populations. Binary interaction processes, including mass transfer (MT), common-envelope (CE) evolution, and tidal interaction, are the primary drivers that distinguish the evolution of binary stars from that of single stars. CE evolution is widely considered to play a fundamental role in the formation of short-period binary systems containing at least one compact companion \citep{1976IAUS...73...75P,2013A&ARv..21...59I}. Despite its importance, the CE phase remains one of the most uncertain, highly debated, and actively studied topics in modern binary stellar evolution.

\begin{figure}[ht!]
\centering
\includegraphics[scale=0.6]{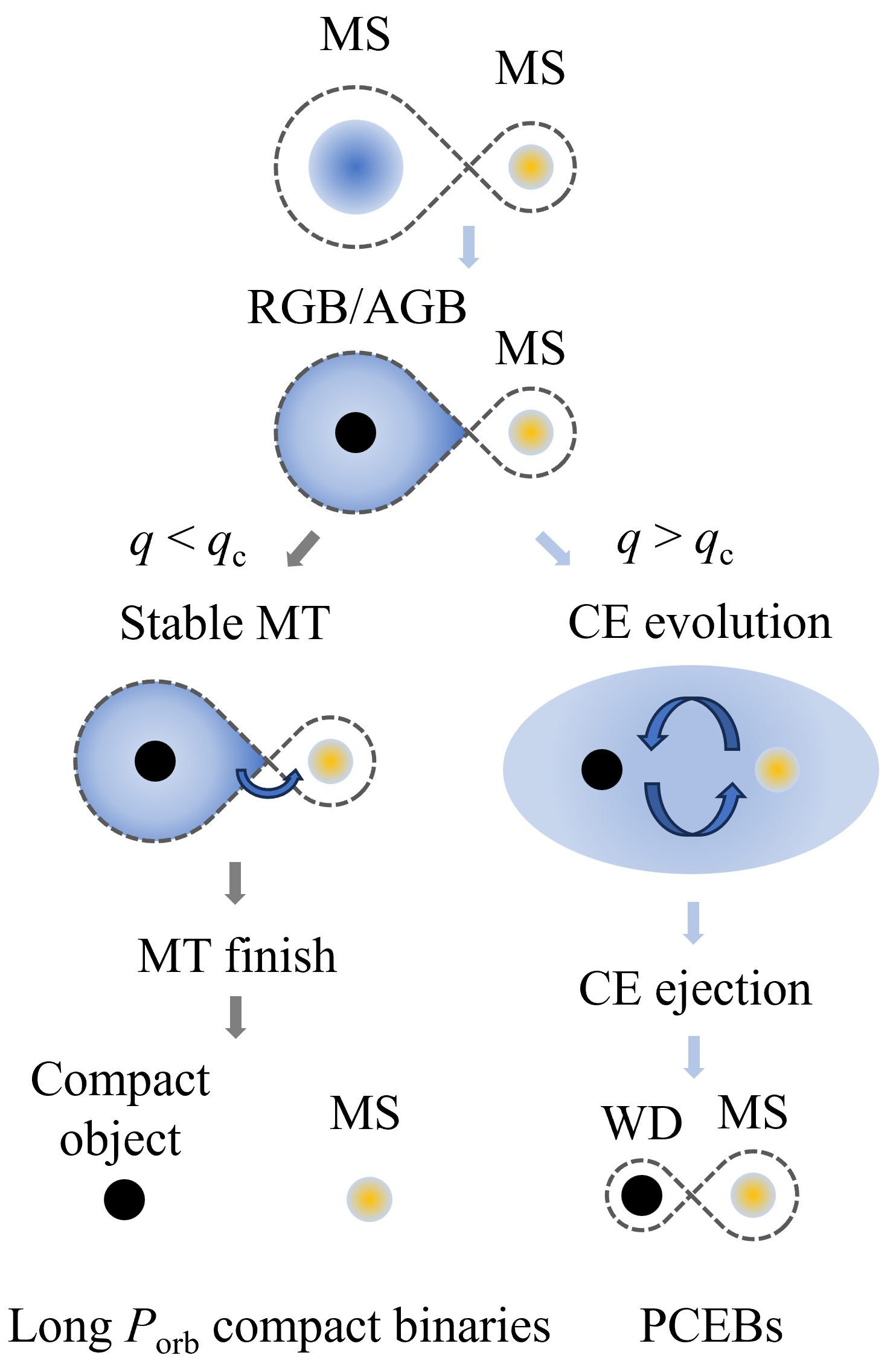}
\caption{Effects of the critical mass ratio ($q_{\rm c}$) and CE formalism on binary evolution and the formation of WD+MS PCEBs. The mass ratio $q$ is defined as the ratio of the donor mass to the accretor mass, and $q_{\rm c}$ denotes the initial maximum $q$ for which stable mass transfer can proceed. Here, we do not consider the grazing envelope evolution (GEE) channel, which may also produce long-period compact binaries (see \autoref{Sec:5.3}).} \label{fig:formation channel}
\end{figure}

Post-common-envelope binaries (PCEBs) are the direct evolutionary products of common-envelope (CE) evolution, and serve as natural astrophysical laboratories, which can be used to constrain the physics of the CE phase and test CE evolution models. Among the diverse population of PCEBs, systems consisting of a compact star and a main-sequence (MS) companion are a particularly well-suited subclass for this purpose. As shown in \autoref{fig:formation channel}, their evolutionary pathways are relatively simple (single formation channel, almost no accretion of matter), and their fundamental binary parameters are predominantly set by the CE phase itself. In this work, we focus exclusively on PCEBs consisting of a white dwarf (WD) and an MS companion that have not undergone a second episode of MT. These systems are primarily formed via low- and intermediate-mass stellar evolution, which is better understood and more theoretically robust than the evolution of massive stars that form neutron stars (NSs) and black holes (BHs). Compared with other PCEBs, they are also sufficiently bright for observational characterization, resulting in a large, abundant sample of systems with well-measured physical parameters.

Over the past two decades, significant advances in our understanding of CE evolution have been made through observational surveys of PCEBs \citep{2011A&A...536A..43N,2011A&A...536A..42Z,2013A&A...557A..87T,2014A&A...566A..86C,2017MNRAS.470.1442C,2024PASP..136h4202Y,2024MNRAS.52711719Y,2025A&A...698A.173T}. In particular, data from the Sloan Digital Sky Survey (SDSS) have yielded a large sample of PCEBs with precisely measured parameters, including WD mass, MS companion mass, and orbital period \citep{2010A&A...520A..86Z,2011A&A...536A..42Z,2011A&A...536A..43N,2012MNRAS.419..806R,2016MNRAS.458.3808R}. These measurements have been widely used to constrain CE ejection efficiency and angular momentum loss mechanisms \citep{2010A&A...513L...7S,2014A&A...566A..86C,2017MNRAS.470.1442C}. The known PCEB population is dominated by systems consisting of a WD with a low-mass MS companion (typically an M dwarf), with orbital periods typically shorter than 10 days; these systems are generally interpreted as the product of inefficient CE evolution \citep{2010A&A...520A..86Z}. There is a sharp decline in the number of systems with companion masses above $0.35\,M_\odot$, a feature widely attributed to disrupted magnetic braking (MB, the quenching of MB in fully convective MS stars, which occurs for stellar masses $M<0.35\,M_\odot$) \citep{2010A&A...513L...7S,2024A&A...682A..33B,2024PASP..136l4201B,2026PASP..138c4202S}.

Previous population synthesis studies of post-CE WD+MS systems have generally adopted MT stability criteria based on polytropic stellar models \citep{1987ApJ...318..794H,2002MNRAS.329..897H}. These models uniformly predict a large population of systems with higher-mass MS companions \citep{2010MNRAS.403..179D,2013A&A...557A..87T,2014A&A...568A..68Z,2016ApJ...826...53A}, but very few such systems have been observed. This dearth of solar-type companion systems is most commonly attributed to disrupted MB \citep{2010A&A...513L...7S,2014A&A...568A..68Z,2024A&A...682A..33B,2024PASP..136l4201B,2026PASP..138c4202S} and observational selection effects \citep{2010MNRAS.403..179D,2012MNRAS.419..806R,2016MNRAS.463.2125P,2023MNRAS.521.1880B}. However, even dedicated surveys specifically targeting PCEBs with A-, F-, G-, or K-type secondary stars \citep{2016MNRAS.463.2125P,2017MNRAS.472.4193R,2020ApJ...905...38R,2021MNRAS.501.1677H,2022MNRAS.512.1843H,2022MNRAS.517.2867H} have not detected a sufficient number of such systems to match the sample size of PCEBs hosting M dwarf companions.

In this paper, we compile the most comprehensive published sample of classic post-CE WD+MS binaries with well-measured WD masses, MS companion masses, and orbital periods. This sample spans both compact and wide systems, as well as both M dwarf and A-/F-/G-/K-type secondaries. By implementing an updated MT stability criterion \citep{2010ApJ...717..724G,2015ApJ...812...40G,2020ApJ...899..132G,2023ApJ...945....7G} and calculation of the envelope structure parameter \citep{2010ApJ...716..114X,2010ApJ...722.1985X}, we investigate whether a single binary population synthesis (BPS) model can self-consistently reproduce the observed parameter distribution of the full PCEB sample, without ad hoc parameter adjustments for specific subpopulations. We find that enhanced mass-transfer stability is an additional mechanism responsible for the observed dearth of PCEBs with solar-type MS companions; neither MB nor selection effects alone can fully account for this deficit, and a combination of all three processes is most likely required.

\section{Binary population synthesis model}\label{Sec:2}

\setcounter{footnote}{0}

We employed the latest version of the rapid binary evolution code binary star evolution (BSE)\footnote{Available at \url{https://astronomy.swin.edu.au/~jhurley/bsedload.html}} \citep{2000MNRAS.315..543H,2002MNRAS.329..897H} with several updates to generate our PCEB population. In brief, we first generate a zero-age MS (ZAMS) population via Monte Carlo sampling based on initial distributions. These systems are then evolved until they reach the maximum evolutionary time. Finally, we identify systems that have formed PCEBs and output the physical parameters of these PCEBs at the maximum evolutionary time. The updates and input parameters are described below. All other parameters not specifically mentioned are set to default in BSE.

\subsection{Initial Binary Parameters}\label{Sec:2.1}

The primary mass is given by the initial mass function (IMF) described by \citet{1979ApJS...41..513M} and \citet{1989ApJ...347..998E},
\begin{equation}
    M_{\rm 1,i} = \dfrac{0.19X}{(1 - X)^{0.75} + 0.032(1 - X)^{0.25}},
    \label{eq:initial_mass}
\end{equation}
where $X$ is a random number uniformly distributed between 0 and 1, and the adopted mass range is $0.1\,M_\odot$ to $100\,M_\odot$. To focus on stars that form WDs and to efficiently utilize our simulation results, we artificially restrict the primary mass range to $0.8\,M_\odot$ to $10\,M_\odot$.

For the secondary, we assumed a flat initial mass ratio distribution,
\begin{equation}
     n(1/q) = \rm constant,
     \label{eq:initial mass ratio}
\end{equation}
where $q=M_{\rm 1}/M_{\rm 2}$ and $0 \le 1/q \le 1$. We set a minimum secondary mass of 
$0.1\,M_\odot$, as brown dwarfs are not considered in this work.

The initial orbital separation \textit{a} follows the distribution \citep{2020RAA....20..161H},
\begin{equation}
    an(a) =
    \begin{cases}
        0.07 \left( \dfrac{a}{a_0} \right)^{1.2}, & a < a_0 \\
        0.07, & a_0 < a < a_1
    \end{cases}
    \label{eq:inital separation}
\end{equation}
where $a_0=10R_\odot$, $a_1=5.75\times10^6 R_\odot$. The distribution implies approximately half of binaries have an orbital period shorter than 100 yr.

We assign a "born time" ($t_{\rm born}$) to each initial binary in our simulation (as in \citealt{2013A&A...549A..95Z}), which denotes the formation epoch of the system, with the formation time of the Milky Way set as the zero-point. The born time equally distributes between 0 and 13 Gyr, and the maximum evolutionary time ($t_{\rm max}$) is naturally defined as $13\,{\rm Gyr} - t_{\rm born}$, corresponding to the assumption of a constant star formation rate.

Lastly, we adopt a metallicity of 0.02, set the initial orbital eccentricity to 0, and initialize the simulation with $10^7$ binaries. Note that these initial conditions are general and do not correspond to a particular Galactic model.

\subsection{Angular Momentum Loss from Magnetic Braking}\label{Sec:2.2}

The MB in BSE takes effect for stars that have appreciable convective envelopes but not fully convective MS stars, and removes a fraction of the orbital angular momentum. The prescription is given by \citet{2002MNRAS.329..897H}, scaled by a normalization factor proposed by \citet{2008MNRAS.389.1563D},
\begin{equation}
\begin{split}
    \dot{J}_{\rm MB} &= - \eta \times 5.83\times10^{-16} \frac{M_{\rm env}}{M_2} 
    \left( \dfrac{R_2}{R_\odot} \right)^3 \\
    &\times \left( \dfrac{\Omega}{\rm rad/ yr} \right)^3 M_\odot R_\odot^2 /{\rm yr^2},
\end{split}
\label{eq:Rap MB}
\end{equation}
where $\eta=0.19$, $M_{\rm env}$ is the convective envelope mass of the secondary star, $R_2$ is the radius of the secondary star, and $\Omega$ is the secondary star’s spin frequency.

\subsection{Mass-transfer Stability Criterion}\label{Sec:2.3}

The critical mass ratio ($q_{\rm c}$) is the key parameter determining the stability of MT, and it also significantly affects the observed PCEB parameter distribution (as shown in \autoref{fig:formation channel}). We adopted the form of fitting formula for $q_{\rm c}$ from \citet{2023ApJ...945....7G} as the physical inputs for rapid binary evolution in BPS studies. The formula of $q_{\rm c}$ is derived from realistic stellar structure models, assuming completely conservative MT in a dynamical timescale.

We perform linear interpolation in the radial range for donors in the very late Hertzsprung gap (HG) to very early red giant branch (RGB) regime, a range not covered by the original fitting formula. \citet{2020ApJS..249....9G} found that if MT initiates on late RGB or the asymptotic giant branch (AGB) phase, binaries with an initial mass ratio greater than $q_{\rm L_2}$ (the $q_{\rm c}$ for the overflow through the outer Lagrangian point ($\rm L_2$) to form CE) would undergo unstable MT via RLOF through $\rm L_2$, which may also result in CE evolution. This implies that a binary with $q_{\rm L_2}$ less than $q_{\rm ad}$ (the $q_{\rm c}$ for dynamical timescale MT derived from the adiabatic mass-loss model) could enter CE evolution before dynamical MT. We therefore set an upper limit of 2 for $q_{\rm c}$ for RGB/AGB donors, which is representative of the typical $q_{\rm L_2}$ values (see Fig.\,9 in \citealt{2020ApJS..249....9G}).

\begin{figure*}[ht!]
\centering
\includegraphics[scale=0.5]{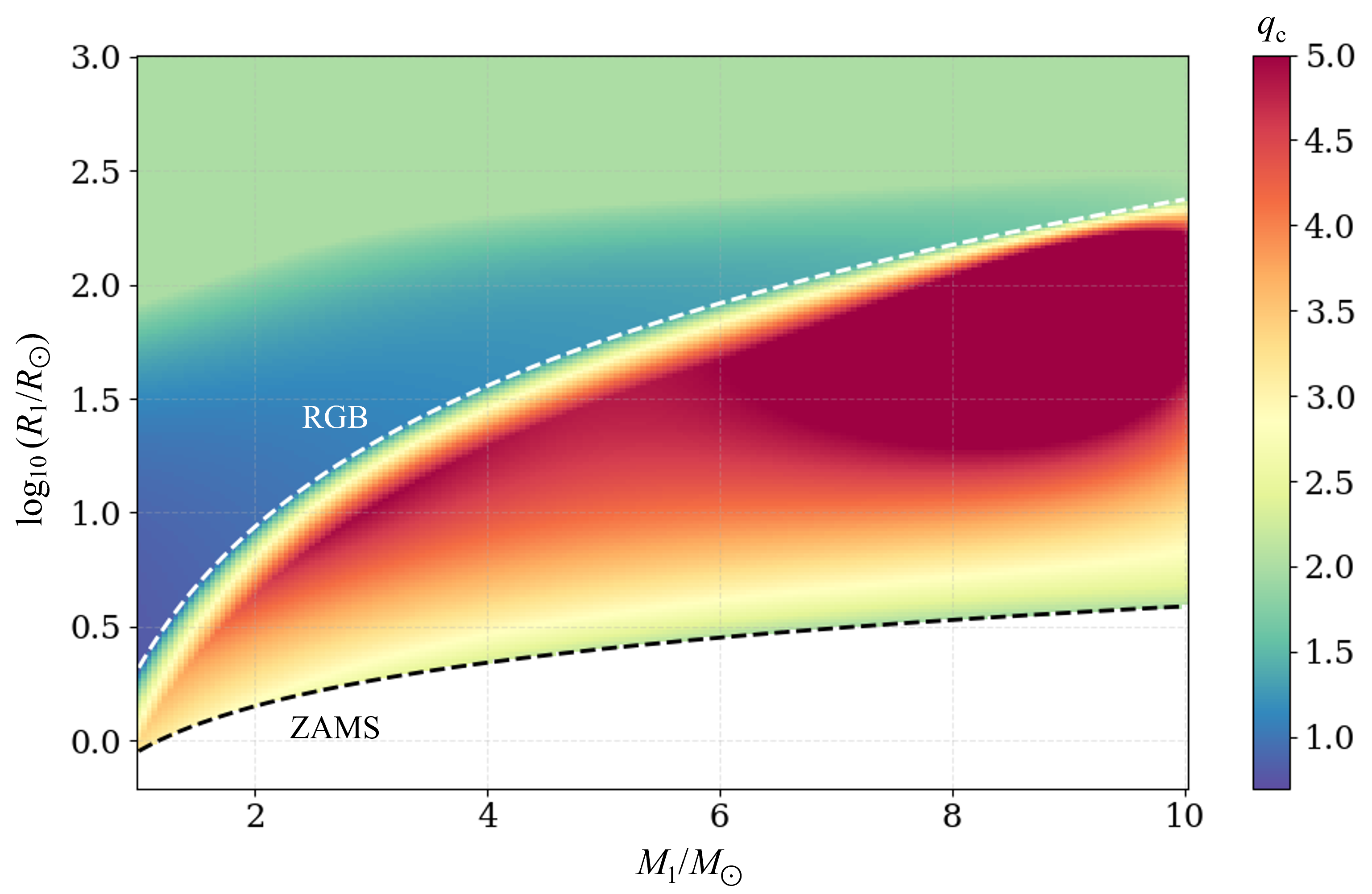}
\caption{The critical mass ratio adopted in our simulations, plotted on the stellar mass-radius diagram. Black-white-dashed lines represent the radii of ZAMS stars and the minimum radii of RGB at different masses, respectively. Note that not all regions shown are physically valid, as the AGB boundary is not plotted here.\label{fig:qc}}
\end{figure*}

The distribution of the adjusted $q_{\rm c}$ on the mass-radius plane is shown in the \autoref{fig:qc}. Compared with the $q_{\rm c}$ from polytropic stellar models, the model of \citet{2020ApJ...899..132G,2023ApJ...945....7G} predicts a higher $q_{\rm c}$ for RGB/AGB donors (see Fig.\,1 in \citealt{2023A&A...669A..82L}). This higher $q_{\rm c}$ makes MT more stable and reduces the number of systems with comparable primary and secondary masses that undergo CE evolution. In other words, for a given primary mass, systems with solar-type secondary companions will avoid CE evolution. We discuss this result in further detail in \autoref{Sec:4}.

\subsection{Common-envelope Evolution}\label{Sec:2.4}

If the mass ratio exceeds $q_{\rm c}$ at the onset of MT, the binary system will initiate CE evolution. The CE phase in BSE is described by the energy conservation formalism \citep{1984ApJ...277..355W,1988ApJ...329..764L}, which links orbital energy loss to the energy required to unbind the donor’s envelope, given by \autoref{eq:CE}.

\begin{equation}
    \alpha_{\rm CE} \left(\frac{{\rm G} M_{\rm 1i} M_2}{2a_{\rm i}}-\frac{{\rm G} M_{\rm 1c} M_2}{2a_{\rm f}} \right)=E_{\rm bind}.
    \label{eq:CE}
\end{equation}

The right-hand side of the equation is the envelope binding-energy term, and the left-hand side is the orbital energy change during the CE phase, where $G$ is the gravitational constant, $M_2$ is the secondary mass, $M_{\rm 1i}$ is the donor mass before the CE phase, $M_{\rm 1c}$ is the core mass of the donor, and $a_{\rm f}$ and $a_{\rm i}$ are the final and initial orbital separations, respectively. $\alpha_{\rm CE}$ is the fraction of orbital energy lost that is used to unbind the CE. The total binding energy $E_{\rm bind}$ of the envelope can be expressed as (e.g., \citealt{2010ApJ...716..114X}),

\begin{equation}
    E_{\rm bind} = \int_{M_{\rm 1c}}^{M_{\rm 1i}} \left( -\dfrac{{\rm G} m(r)}{r} + \alpha_{\rm th} U \right) {\rm d}m,
    \label{eq:binding_energy}
\end{equation}

The first and the second terms in the integral correspond to the gravitational and internal energy of stellar matter, respectively, with $\alpha_{\rm th}$ being the percentage of internal energy contributing to envelope ejection.

For convenience, the total binding energy is usually expressed as
\begin{equation}
    E_{\rm bind} = -\dfrac{{\rm G}M_{\rm 1i}M_{\rm 1e}}{\lambda R_{\rm 1i}},
    \label{eq:binding_energy 2}
\end{equation}
where $M_{\rm 1e} = M_{\rm 1i}-M_{\rm 1c}$ is the mass of the donor’s envelope, $R_{\rm 1i}$ is the donor’s radius before the CE phase, and $\lambda$ is the envelope structure parameter reflecting the structure of the donor.

We use the "Nanjing lambda" fitting formula for $\lambda$ from \citet{2010ApJ...716..114X,2010ApJ...722.1985X}, which is calibrated against detailed stellar evolution models. They calculated $\lambda$ values for donor masses from $1\,M_\odot$ to $20\,M_\odot$ across a range of radii and evolutionary phases, and defined two $\lambda$ prescriptions: $\lambda_{\rm g}$ (with $\alpha_{\rm th}=0$), which only includes the contribution of gravitational binding energy, and $\lambda_{\rm b}$ (with $\alpha_{\rm th}=1$), which also includes the contribution of full internal energy (thermal energy, radiation energy, full ionization energy, and the dissociation energy of molecular hydrogen). 

We perform linear interpolation between the mass anchor points of the original formula to fill in the uncalibrated mass ranges. To enable consistent interpolation across all masses, we refit $\lambda_{\rm g}$ and $\lambda_{\rm b}$ as functions of stellar radius for the $1\,M_\odot$ track, replacing the default parameterization in terms of envelope mass fraction. Furthermore, if MT initiates during the late thermally pulsing AGB (TP-AGB) phase, the donor radius often exceeds the maximum radius covered by the fitting formula. In this case, we adopt constant extrapolation using the last valid calculated value of $\lambda_{\rm g}$ and $\lambda_{\rm b}$. For $0.986\,M\odot < M_{\rm 1,ZAMS} < 1\,M_\odot$, we adopt the $1\,M_\odot$ fitting formula exclusively; systems with $M_{\rm 1,ZAMS} < 0.986\,M_\odot$ cannot form PCEBs within a 13 Gyr galaxy age. 

\section{Observational Samples}\label{Sec:3}

All PCEB observational samples used in this work are compiled from published papers and preprints. Our sample selection follows two key criteria: 

\textup{(1)} systems robustly classified as WD+MS PCEBs (the compact object is a WD; and the companion is an MS star; with no accretion disk, and the companion has not filled its RL; they do not lie on the $M_{\rm WD}-P_{\rm orb}$ relation for stable MT; typically with long WD cooling ages and circular orbits; are unlikely to be detached cataclysmic variables, hereafter CVs, within the period gap) that have not undergone a second MT

\textup{(2)} and systems with well-measured simultaneous values of the WD mass, MS companion mass, and orbital period, excluding systems with only upper or lower limits on any of these three parameters.

The observed physical properties of the PCEBs in our final sample are summarized in \autoref{tab:observation}.

Our sample is primarily identified from spectroscopic and eclipsing binary surveys using data from the SDSS \citep{2011A&A...536A..42Z,2011A&A...536A..43N} and the Zwicky Transient Facility (ZTF; \citealt{2023MNRAS.521.1880B,2026PASP..138c4202S}). The original WD masses derived from SDSS spectral fits for multiple systems in the \citet{2011A&A...536A..43N} sample exhibit significant biases due to the log g upturn problem, especially for massive WDs. We therefore adopt corrected WD masses for these affected systems in this sample, which were recalculated specifically for this work by applying the tabulated 3D corrections of \citet{2013A&A...559A.104T}. The corrected values are listed in \autoref{tab:Corrected M_WD}.

\section{Results}\label{Sec:4}

To quantify the impact of our adopted new MT stability criterion, we perform BPS simulations for six distinct models. These models combine two prescriptions for the critical mass ratio $q_{\rm c}$ and three values of the CE efficiency parameter $\alpha_{\rm CE}$. In these models, we only consider the gravitational energy of the envelope with no additional energy contributions ($\alpha_{\rm th}=0$). The impact of including internal energy is discussed in the following section. The model parameters and key simulation outputs are summarized in \autoref{tab:N_PCEB}

\begin{deluxetable}{cccc}[ht!]
\tabletypesize{\scriptsize}
\tablecaption{Model parameters and key simulation outputs for the six models adopted in this work.\label{tab:N_PCEB}}
\tablehead{
\colhead{Model} & \colhead{$q_{\rm c}$} & \colhead{$\alpha_{\rm CE}$} & \colhead{$N_{\rm PCEB}$}
}
\startdata
a & Ge   et al & 0.25 & 45,629   \\
b & Ge   et al & 0.5  & 114,298  \\
c & Ge   et al & 1.0  & 217,199  \\
d & Polytropic & 0.25 & 90,891   \\
e & Polytropic & 0.5  & 194,268  \\
f & Polytropic & 1.0  & 329,257 
\enddata
\tablecomments{$N_{\rm PCEB}$ is the total number of simulated PCEBs produced by each model.}
\end{deluxetable}

\subsection{Simulation Results}\label{Sec:4.1}

\begin{figure*}[p!]
\centering
\includegraphics[scale=0.57]{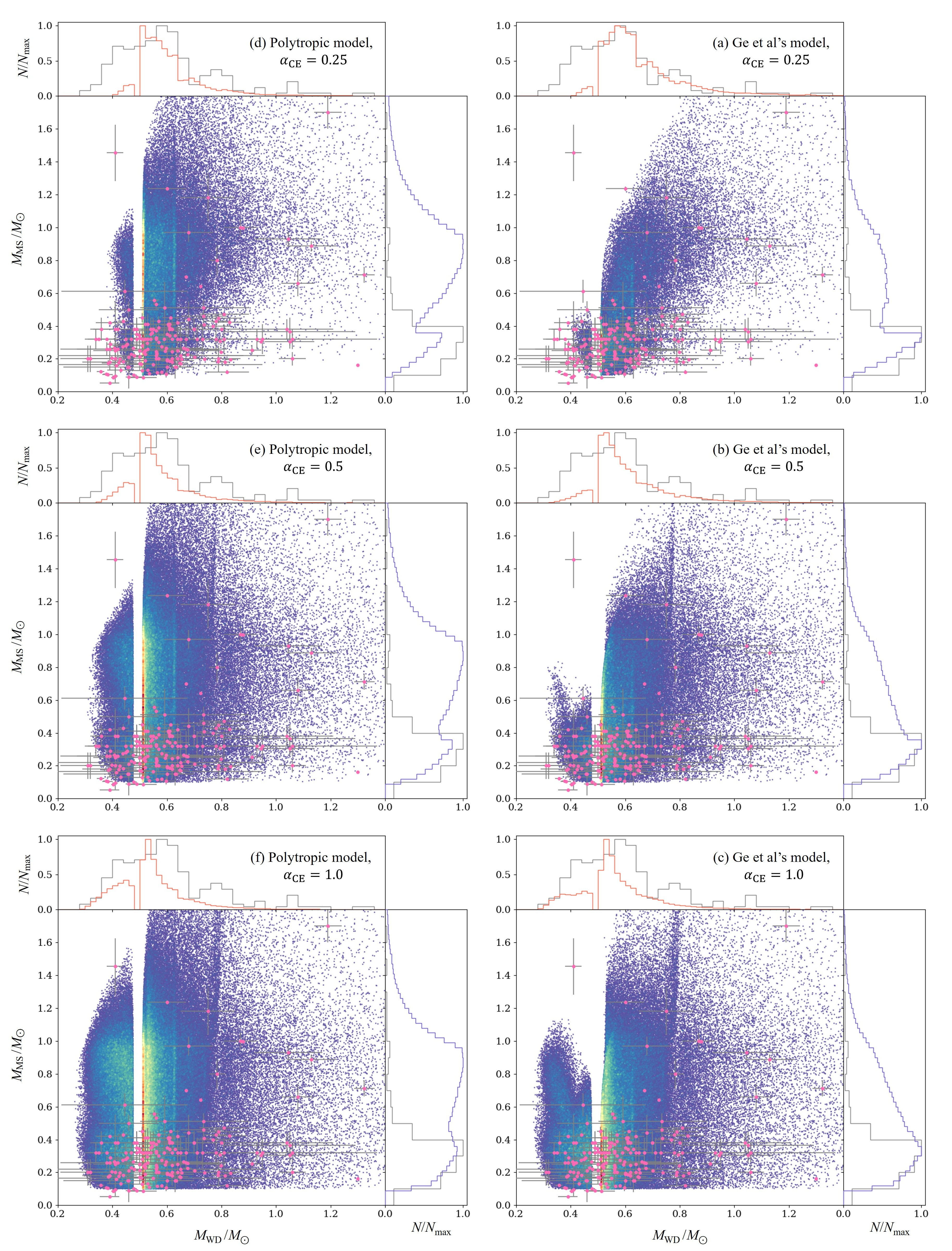}
\caption{The distribution of observed (pink dots) and simulated PCEBs in the $M_{\rm WD}-M_{\rm MS}$ plane. The point color encodes the local number density of simulated PCEBs, normalized independently to the maximum density in each panel for visual clarity. Peak-normalized marginal histograms are shown at the top ($M_{\rm WD}$) and right ($M_{\rm MS}$), where gray lines represent the observed distributions, red represents the simulated $M_{\rm WD}$ distribution, and blue represents the simulated $M_{\rm MS}$ distribution. Gray error bars represent the 1$\sigma$ uncertainties in both coordinates for each observed system.\label{fig:M_WD-M_MS}}
\end{figure*}

\begin{figure*}[p!]
\centering
\includegraphics[scale=0.57]{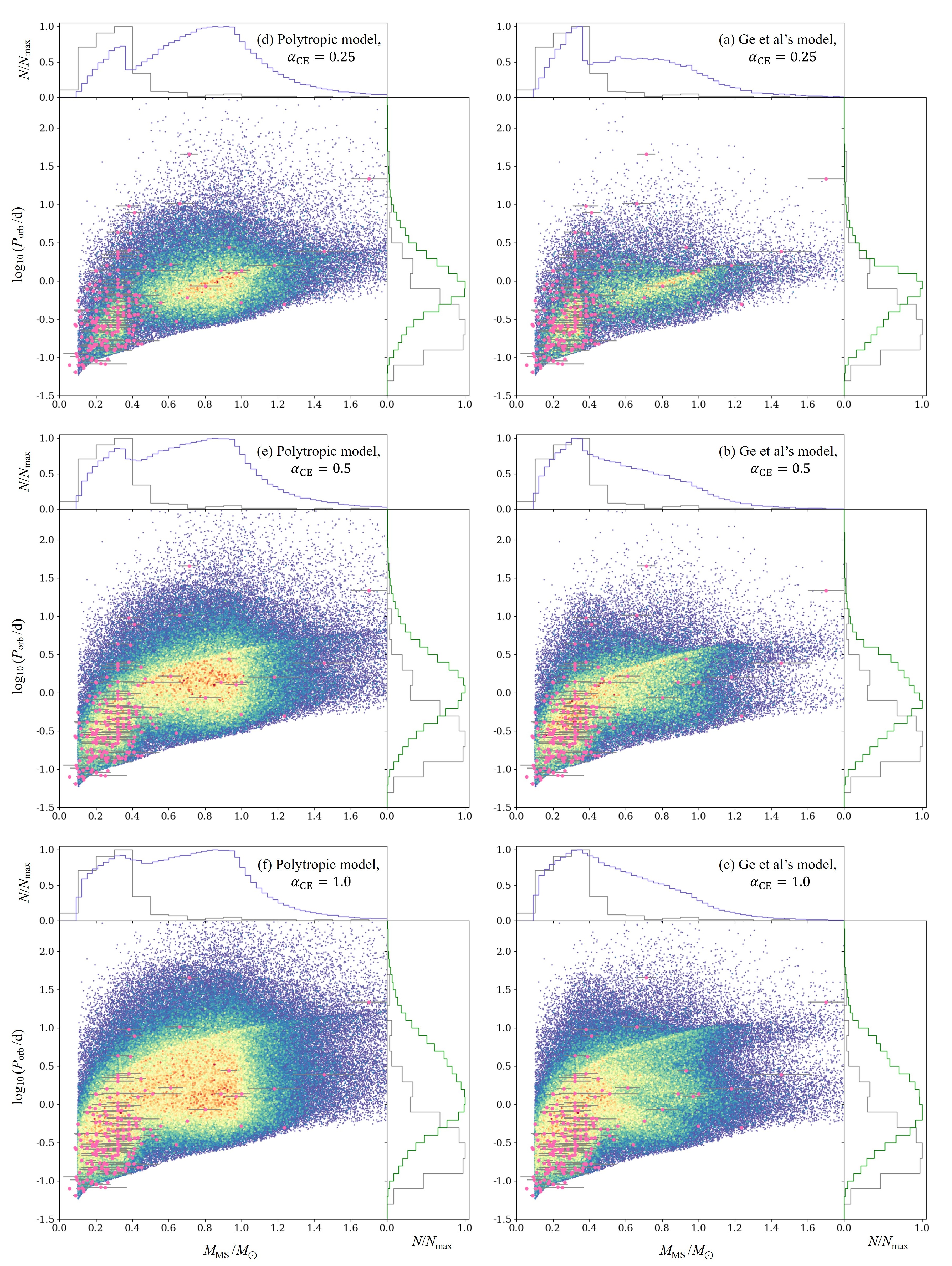}
\caption{The distribution of observed (pink dots) and simulated PCEBs in the $M_{\rm MS}-P_{\rm orb}$ plane. The color intensity scale is identical to \autoref{fig:M_WD-M_MS}. Right green marginal histograms show the peak-normalized distribution of ${\rm log}_{10}(P_{\rm orb})$. Note that orbital period measurements are sufficiently precise; we do not show their error bars. This does not imply that points without error bars in $M_{\rm WD}$ or $M_{\rm MS}$ are perfectly measured; it only means that no definitive upper or lower limits are available for those parameters.\label{fig:M_MS-P_orb}}
\end{figure*}

\begin{figure*}[p!]
\centering
\includegraphics[scale=0.57]{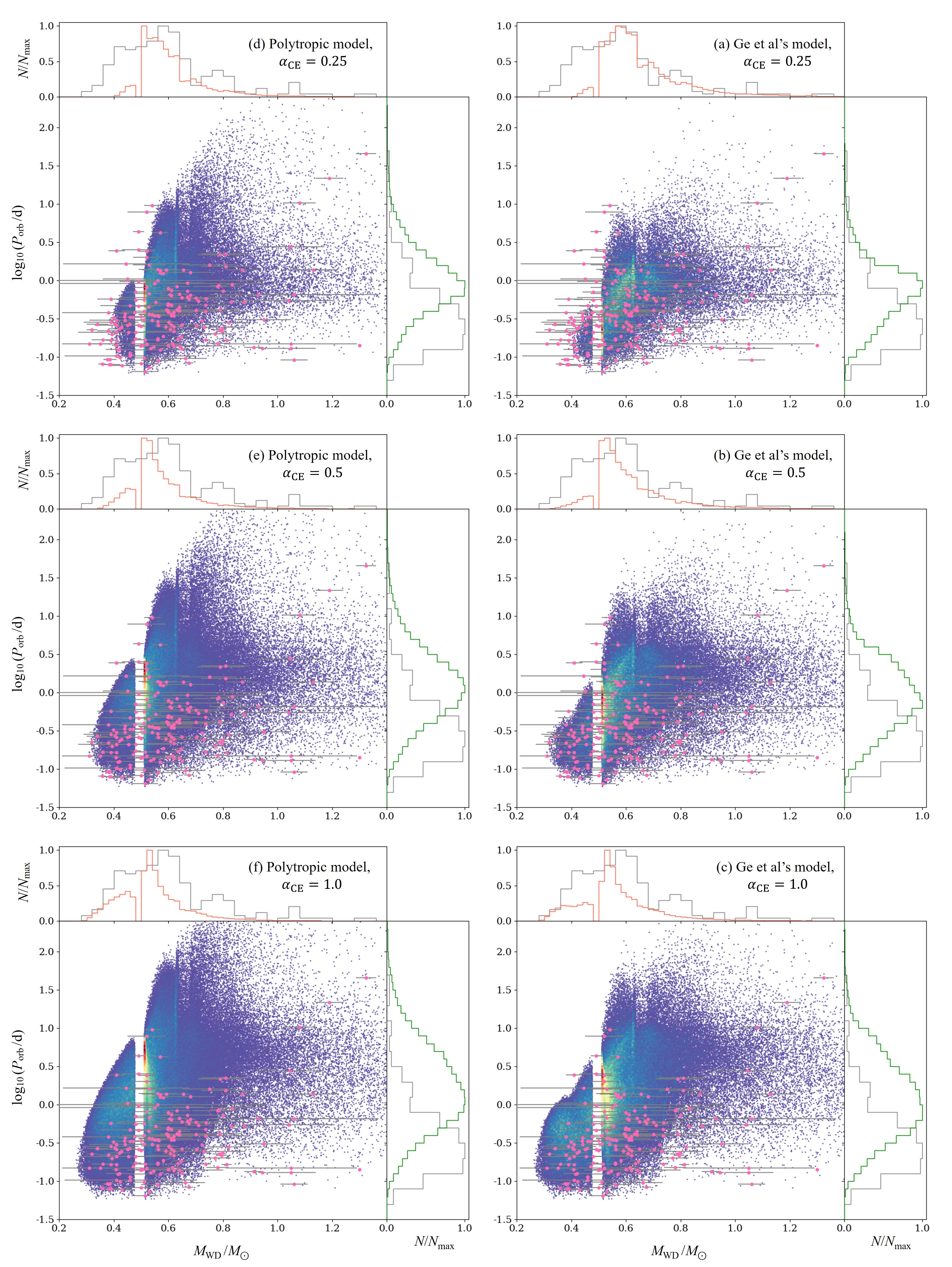}
\caption{The distribution of observed (pink dots) and simulated PCEBs in the $M_{\rm WD}-P_{\rm orb}$ plane. The color intensity scale is identical to Figures \ref{fig:M_WD-M_MS} and \ref{fig:M_MS-P_orb}. \label{fig:M_WD-P_orb}}
\end{figure*}

Figures \ref{fig:M_WD-M_MS}, \ref{fig:M_MS-P_orb} and \ref{fig:M_WD-P_orb} show the distribution of simulated PCEBs from all six models in the $M_{\rm WD}-M_{\rm MS}$, $M_{\rm MS}-P_{\rm orb}$, and $M_{\rm WD}-P_{\rm orb}$ planes, respectively. Compared with the polytropic $q_{\rm c}$ model, the most prominent feature of the new criterion \citep{2023ApJ...945....7G} is the dearth of PCEBs with solar-type MS companions, the very systems whose absence drives the lower total PCEB counts for the H. Ge et al.'s models listed in \autoref{tab:N_PCEB}. 

\begin{figure}[ht!]
\centering
\includegraphics[scale=0.65]{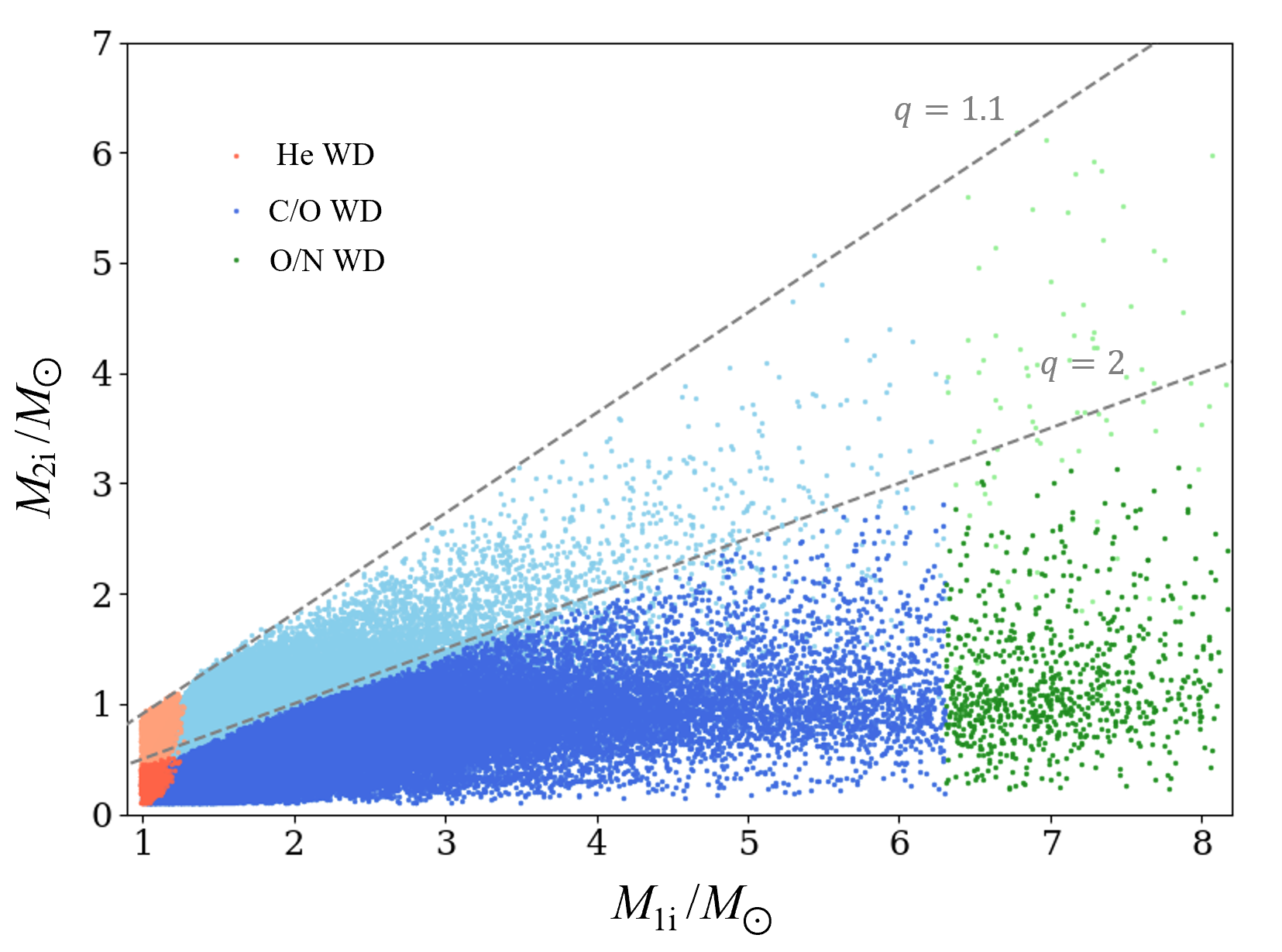}
\caption{The initial primary versus initial secondary mass distribution of PCEB progenitors. Points are color-coded by the final WD type: red for He WDs, blue for carbon--oxygen (C/O) WDs, and green for oxygen--neon (O/Ne) WDs. Dark points in the foreground correspond to the H. Ge et al. $q_{\rm c}$ prescription (model (a)), and are overlaid on top of lighter points representing the polytropic $q_{\rm c}$ prescription (model (d)). The gray-dashed lines mark constant mass ratios of $q=1.1$ and $q=2$, providing a direct visual reference for the two different $q_{\rm c}$ thresholds.\label{fig:M1i-M2i}}
\end{figure}

To explain this difference, we show the initial mass distribution of PCEB progenitors for models (a) and (d) in \autoref{fig:M1i-M2i}. It is clear that H. Ge et al.'s $q_{\rm c}$ prescription imposes a much higher minimum mass ratio ($\sim2$) for entering CE evolution than the polytropic $q_{\rm c}$ prescription ($\sim1.1$). This directly prevents binary systems with comparable primary and secondary masses from entering the CE phase and significantly reduces the number of systems with solar-type secondaries that form PCEBs. For the progenitors of helium (He) WDs, the initial stellar mass is typically lower ($<1.6\,M_\odot$) than that of carbon--oxygen (C/O) WD progenitors; the maximum allowed companion mass is therefore also lower. This explains the clear deficit of PCEBs with He WDs and solar-type companions in the panels using H. Ge et al.'s $q_{\rm c}$ prescription in \autoref{fig:M_WD-M_MS}. Besides, one may wonder why the minimum mass ratio for the polytropic $q_{\rm c}$ prescription is $\sim1.1$ rather than 1. Although the polytropic $q_{\rm c}$ prescription even allows systems with mass ratios below 1 to enter CE evolution, for binaries with excessively similar initial masses, the secondary will evolve off the MS and ascend the giant branch before the primary initiates CE evolution and forms a WD, thereby precluding the formation of WD+MS PCEBs.

The full evolutionary track of a representative binary system with comparable primary and secondary masses, evolved under both $q_{\rm c}$ prescriptions and presented in Tables \ref{tab:default_qc} and \ref{tab:ge_qc}, illustrates this difference more intuitively. For the default BSE $q_{\rm c}$ prescription \citep{2002MNRAS.329..897H}, the donor initiates Roche lobe (RL) overflow (RLOF) on the RGB when $M_1/M_2 > q_{\rm c}$ (0.83), triggering CE evolution and the formation of a short-period He WD PCEB. In contrast, with the new $q_{\rm c}$ prescription \citep{2023ApJ...945....7G}, the mass ratio $M_1/M_2 < q_{\rm c}$ (2.00) at the onset of RLOF. The system therefore undergoes stable MT instead of CE evolution, and the donor eventually evolves into a C/O WD with a massive MS companion in a wide orbit.

In addition, we find that the width of the WD mass distribution broadens, and the number of simulated PCEBs (\autoref{tab:N_PCEB}) increases monotonically with increasing $\alpha_{\rm CE}$. This is because increasing $\alpha_{\rm CE}$ significantly enhances orbital energy utilization efficiency, particularly for systems hosting low-mass He WDs and low-mass MS companions. This allows these systems to either survive the CE phase or retain longer orbital periods after CE ejection; they therefore remain in the PCEB phase for a longer duration before evolving into CVs.

We further find that the suppressive effect of the H. Ge et al. $q_{\rm c}$ prescription on PCEBs with solar-type MS companions appears to weaken as $\alpha_{\rm CE}$ increases. This arises from two complementary physical effects. First, as noted above, higher $\alpha_{\rm CE}$ enables a large population of low-mass He WD systems to survive the CE phase. These systems form via MT initiated in the early RGB phase, when the stellar radius is still small, and the corresponding $q_{\rm c}$ is low ($\sim1.1$, see \autoref{fig:qc}). This naturally permits systems with more massive companions to enter the CE phase, resulting in a distinct upward, leftward extending tail in the He WD distribution for models adopting the H. Ge et al. criterion as $\alpha_{\rm CE}$ increases (see \autoref{fig:M_WD-M_MS}). Second, increasing $\alpha_{\rm CE}$ shifts the entire PCEB population toward longer orbital periods. MB, however, only induces significant orbital decay for short-period PCEBs, driving them to rapidly evolve into CVs. A systematic shift to longer periods means far fewer systems are strongly affected by MB, which explains why the prominent dip in the companion mass distribution around $0.35\,M_\odot$ is clearly visible in lower $\alpha_{\rm CE}$ models but absent in the $\alpha_{\rm CE}=0.5$ and $\alpha_{\rm CE}=1.0$ models in \autoref{fig:M_MS-P_orb}.

In Figures \ref{fig:M_WD-M_MS} and \ref{fig:M_WD-P_orb}, we identify a prominent common feature: a clear gap in WD mass around $0.5\,M_\odot$. The left edge of this WD mass gap corresponds to the maximum He core mass for stars ascending at the RGB. If a star evolves to the tip of the RGB (TRGB) without filling its RL, it ignites core He burning, shrinks in radius, and evolves to the horizontal branch. When the star later enters the AGB and expands again, its radius remains smaller than its TRGB radius. The core mass therefore continues to grow until the stellar radius exceeds the TRGB radius and fills the RL, eventually forming a C/O WD on the high-mass side of the gap.

\subsection{Observational Verification}\label{Sec:4.2}

To validate our models, we overplot the observational sample presented in Section \ref{Sec:3} onto the simulated PCEB distributions in the $M_{\rm WD}-M_{\rm MS}$, $M_{\rm MS}-P_{\rm orb}$, and $M_{\rm WD}-P_{\rm orb}$ planes, as shown in Figures \ref{fig:M_WD-M_MS}, \ref{fig:M_MS-P_orb}, and \ref{fig:M_WD-P_orb}, respectively. We find that models adopting the polytropic $q_{\rm c}$ prescription significantly overpredict the number of PCEBs with solar-type MS companions ($>0.4\,M_\odot$) relative to observations. In contrast, models using the \citet{2023ApJ...945....7G}  $q_{\rm c}$ criterion show much better agreement with the observed distribution.

Furthermore, we find that models with larger values of $\alpha_{\rm CE}$ deviate significantly from the observations in both the companion mass distribution and the orbital period distribution. Among all models, the $\alpha_{\rm CE}=0.25$ model provides the best overall reproduction of the observed parameter space. This is consistent with previous studies concluding that the majority of short-period PCEBs form via inefficient CE evolution \citep{2010A&A...520A..86Z,2013A&A...557A..87T,2014A&A...566A..86C,2017MNRAS.470.1442C,2022ApJ...933..137G,2024ApJ...961..202G,2025A&A...695A.161S}.

\section{Discussion}\label{Sec:5}

\subsection{Effect of Stellar Wind on the Calculation of $q_{\rm c}$}\label{Sec:5.1}

In our BPS models, we include the effects of wind mass loss, and calculate the critical mass ratio using the instantaneous mass of the donor at the onset of MT. However, the $q_{\rm c}$ fitting formula is derived from detailed stellar evolution models without accounting for wind mass loss. \citet{2020ApJ...899..132G} suggested that the $q_{\rm c}$ for giant stars with wind mass loss should instead adopt the value for a wind-free star with the same ZAMS mass and core mass. Nevertheless, this inconsistency has a negligible impact on our results. By the time stellar wind causes significant mass loss relative to the star’s ZAMS mass, the donor has already evolved to the late RGB or AGB phase, where $q_{\rm c}$ is already set to our constant upper limit of 2 (see \autoref{Sec:2.3}). Thus, $q_{\rm c}$ remains fixed at 2 regardless of whether we calculate it from the instantaneous donor mass or the original ZAMS mass.

\subsection{Effect of the Magnetic Braking Prescription}\label{Sec:5.2}

In \autoref{Sec:2.2}, we adopt a weaker MB prescription than the default formalism in the BSE code. However, recent studies \citep{2023A&A...678A..34B,2023MNRAS.520.3187S,2025A&A...703A.119T} suggest that only a stronger MB prescription can reproduce the formation of AM CVn systems via the CV channel. We therefore investigate whether a stronger MB prescription can explain the observed dearth of PCEBs with solar-type companions. Given the effective range of MB, we apply the stronger MB prescription only to MS stars to focus on post-CE evolution, and retain the default prescription for all other evolutionary stages.

For MS stars with a substantial convective envelope, we adopt the convection and rotation boosted (CARB) prescription from \citet{2019ApJ...886L..31V}, given by

\begin{equation}
\begin{split}
\dot{J}_{\rm MB} &= -2 \times 10^{-6} 
\left( \dfrac{-\dot{M}_{\rm wind}}{\rm g/s} \right)^{-1/3}
\left( \dfrac{R}{\rm cm} \right)^{14/3}\\
&\quad \times \left( \dfrac{\Omega}{\Omega_\odot} \right)^{11/3}
\left( \dfrac{\tau_{\rm conv}}{\tau_{\odot,{\rm conv}}} \right)^{8/3}
\left[ \left( \dfrac{v_{\rm esc}}{\rm cm/s} \right)^2 \right. \\
&\quad \left. + \dfrac{2}{K_2^2} \left( \dfrac{\Omega}{\rm s^{-1}} \right)^2 \left( \dfrac{R}{\rm cm} \right)^2 \right]^{-2/3}
{\rm g\ cm^2 / s^2},
\end{split}
\label{eq:mb_angular_momentum}
\end{equation}

where $\dot{M}_{\rm wind}$, $R$, $\Omega$, $\tau_{\rm conv}$, $v_{\rm esc}$ are the wind mass-loss rate, radius, spin angular velocity, convective turnover timescale, and surface escape velocity of the star, respectively. $K_2$ is set to a constant value of 0.07, and $\Omega_\odot=94.6$ rad yr$^{-1}$.

Since the original BSE code does not output detailed stellar structure information, we interpolate the convective turnover timescale as a function of the instantaneous MS mass, using the tabulated grids from \citet{2025ApJ...988..102G}. For this calibration, the solar convective turnover timescale $\tau_{\odot,{\rm conv}}$ is set to $\sim$14 days.

\begin{figure*}[ht!]
\centering
\includegraphics[scale=0.58]{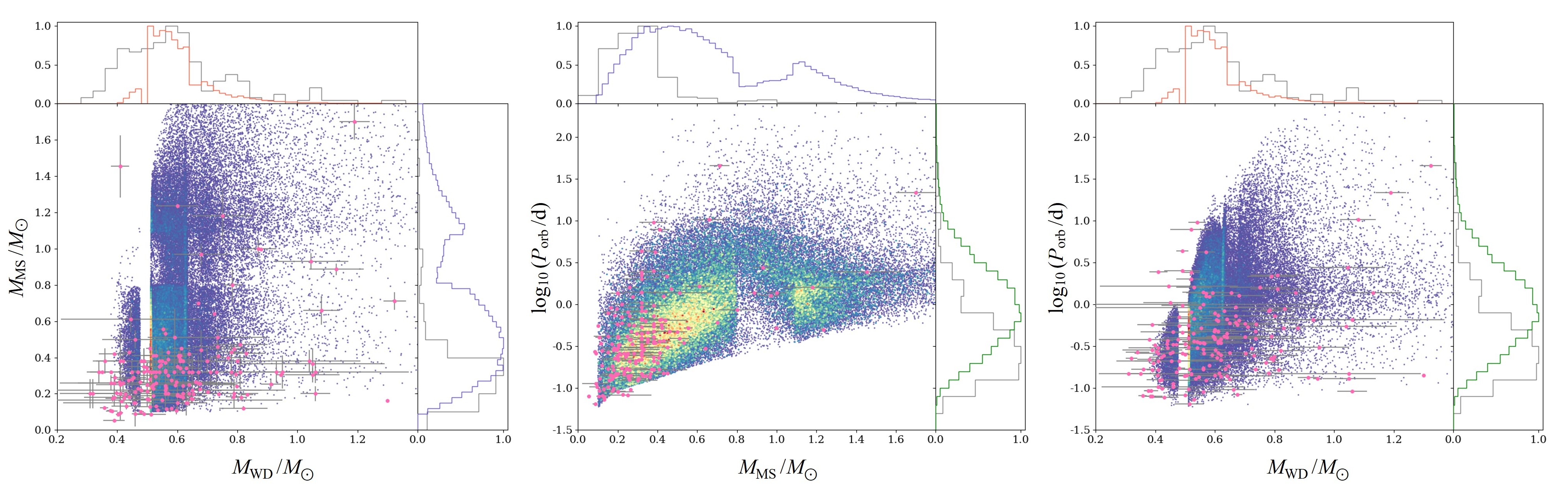}
\caption{The distribution of observed (pink dots) and simulated PCEBs using the CARB MB prescription, plotted in the $M_{\rm WD}-M_{\rm MS}$, $M_{\rm MS}-P_{\rm orb}$, and $M_{\rm WD}-P_{\rm orb}$ planes. The color scheme is identical to \autoref{fig:M_WD-M_MS}.\label{fig:CARB obs}}
\end{figure*}

We retain all other parameters from model (d), only changing the MB prescription to CARB for MS stars. \autoref{fig:CARB obs} shows the distribution of the simulated PCEBs overlaid with the observational sample. While the stronger MB prescription does cause a fraction of systems with solar-type companions to disappear, a large excess of systems with MS companion masses between $0.35\,M_\odot$ and $0.8\,M_\odot$, as well as above $1.1\,M_\odot$, remains. This discrepancy arises for two reasons: (1) For MS stars more massive than $1.1\,M_\odot$, the convective envelope becomes progressively thinner and eventually disappears, so MB ceases to operate entirely. (2) For MS stars less massive than $0.8\,M_\odot$ (but not fully convective, $M>0.35\,M_\odot$), MB is strongly suppressed by the weak tidal coupling strength.

In the standard picture, MB first reduces the stellar spin angular momentum, making the spin angular velocity slower than the orbital angular velocity. Tides then transfer orbital angular momentum to the stellar spin to maintain tidal synchronization, allowing MB to continuously drain orbital angular momentum and shrink the orbit. However, this process becomes ineffective if MB is too strong.

\autoref{fig:MB compare} shows an example of this effect for a PCEB with $M_{\rm WD}=0.5747\,M_\odot$, $M_{\rm MS}=0.3666\,M_\odot$, and a post-CE orbital period of $\sim$1.14 days, evolved under three different MB prescriptions. Tides act to spin down the star when $\Omega > \Omega_{\rm eq}$ (the tidal synchronization angular velocity) and spin it up when $\Omega < \Omega_{\rm eq}$, with the tidal torque reaching its maximum at $\Omega \approx \Omega_{\rm eq}$. This saturation arises because the dominant contributors to the frictional torque are the largest convective cells; when the difference between $\Omega$ and $\Omega_{\rm eq}$ is too large, these cells can no longer effectively contribute to the viscosity (see Section 2.3.1 in \citealt{2002MNRAS.329..897H} for details). Thus, the effectiveness of MB depends critically on whether tides can replenish the spin angular momentum lost to MB while the system remains near tidal synchronization ($\Omega \approx \Omega_{\rm eq}$). For the stronger CARB and default prescriptions, the MB torque remains larger than the tidal torque even at $\Omega \approx \Omega_{\rm eq}$. The stellar spin therefore decreases rapidly, and the tidal torque drops off in turn. Even if the tidal torque eventually matches the MB torque, MB cannot effectively drain orbital angular momentum, as the spin velocity remains too low to sustain tidal coupling. In contrast, the weaker MB prescription adopted in our fiducial model allows tides to match the MB torque, keeping the system tidally locked and the spin velocity sufficiently high for MB to efficiently remove orbital angular momentum via tidal coupling.

This leads to the dramatic result that a stronger MB prescription can produce weaker net orbital angular momentum loss. It is important to note that this effect is not universal for systems with solar-type companions: for these systems, the tidal torque is stronger, and the MB torque is weaker than for low-mass MS companions at the same orbital period, so tides can easily keep up with the spin angular momentum loss from MB. This explains why a fraction of the solar-type companion systems do disappear under the CARB prescription.

Similar results have been found in previous studies \citep{2019ApJ...881...88F,2024ApJ...971...80S}, but a more systematic and comprehensive investigation of tide--MB decoupling under different MB prescriptions and its impact on PCEB evolution is needed. We conclude that, at least within the framework of the default tidal model \citep{1981A&A....99..126H} implemented in the BSE code, attributing the dearth of PCEBs with solar-type companions solely to stronger MB, without accounting for the $q_{\rm c}$ prescription, is not justified, unless we ignore tidal--MB coupling and artificially drain orbital angular momentum directly from the binary orbit.

\subsection{Formation of Long-period WD+MS Binaries}\label{Sec:5.3}

In recent years, a large number of WD+MS binaries with orbital periods ranging from tens to hundreds of days have been identified using data from the Gaia mission \citep{2024PASP..136h4202Y,2024MNRAS.52711719Y,2026AJ....171..159M}. This wide-orbit WD+MS population was not predicted by prior BPS models, and may imply efficient envelope ejection during CE evolution with only limited orbital period decay. To resolve this tension between theory and observations, it has been proposed that additional energy sources, including recombination energy and/or the internal energy of the envelope, contribute to the envelope ejection process \citep{2024A&A...687A..12B,2024A&A...686A..61B,2024PASP..136h4202Y,2024MNRAS.52711719Y}

Here, we also investigate whether including the contribution of internal energy can produce long-period PCEBs in our models. We adopt a linear combination of the $\lambda_{\rm g}$ and $\lambda_{\rm b}$ parameters introduced in Section \ref{Sec:2.4} (as implemented in the COMPAS code; \citealt{2022ApJS..258...34R}), given by $\lambda= \alpha_{\rm th}\lambda_{\rm b} + (1-\alpha_{\rm th})\lambda_{\rm g}$. We take models (a) and (d) as our baseline models and rerun the simulations with only $\alpha_{\rm th}$ set to 0.5, which corresponds to including half of the contribution from internal energy. An important caveat is that including internal energy contributions can cause the total envelope binding energy to become positive at late evolutionary stages. In such cases, the envelope may be ejected via dynamical ejection or superwind \citep{1994MNRAS.270..121H}, and the standard $\alpha_{\rm CE
}$ formalism for CE evolution breaks down \citep{2024PASP..136h4202Y}. Therefore, in our models with internal energy, we only consider systems that initiate MT and form PCEBs before the envelope binding energy becomes positive.

\begin{figure*}[ht!]
\centering
\includegraphics[scale=0.56]{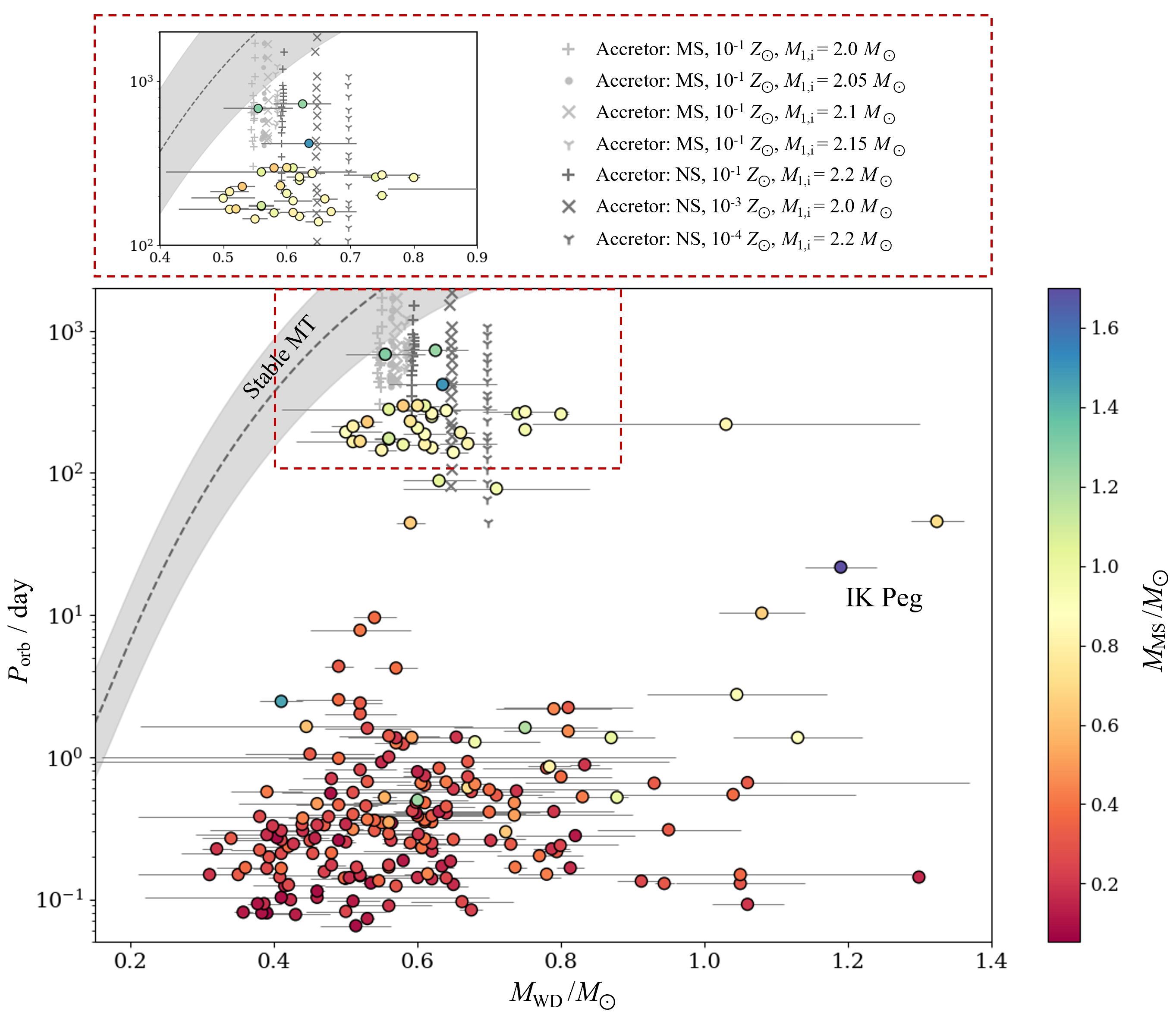}
\caption{Our compiled sample of classic PCEBs and long-period WD+MS binaries (from \citealt{2014Sci...344..275K,2018AJ....155..144K,2024PASP..136h4202Y}) are plotted in the $M_{\rm WD}-P_{\rm orb}$ plane (adapted from \citealt{2024PASP..136h4202Y}; and \citealt{2023MNRAS.518.4579P}). Point colors encode the mass of the MS companion in each system. The location of the well-known PCEB system IK Peg is marked for reference. The dashed line and shaded gray region show the WD mass–final orbital period relation at the end of stable mass transfer (with an orbital period diffusion factor of 2.4) from \citet{1995MNRAS.273..731R}. Light gray plus signs, dots, crosses, and triangles represent results from the stable mass-transfer models of \citet{zheng2026massorbitalperioddistributionmassive}, computed with a fixed initial $M_{\rm MS}=1.4\,M_\odot$ and $Z=0.1\,Z_\odot$. Dark gray plus signs, crosses, and triangles show corresponding results from the same study for systems with a $1.4\,M_\odot$ NS companion.\label{fig:obs fig}}
\end{figure*}

Our long-period WD+MS sample is taken from \citet{2024PASP..136h4202Y} and four self-lensing binaries \citep{2014Sci...344..275K,2018AJ....155..144K}. We plot these systems together with our compiled sample of classic PCEBs in \autoref{fig:obs fig} for comparison. The classic PCEB population typically consists of short-period ($\sim$days) systems with low-mass M dwarf companions (red points in \autoref{fig:obs fig}), while the long-period WD+MS systems have orbital periods of up to hundreds of days and more massive companions.

\begin{figure*}[p!]
\centering
\includegraphics[scale=0.57]{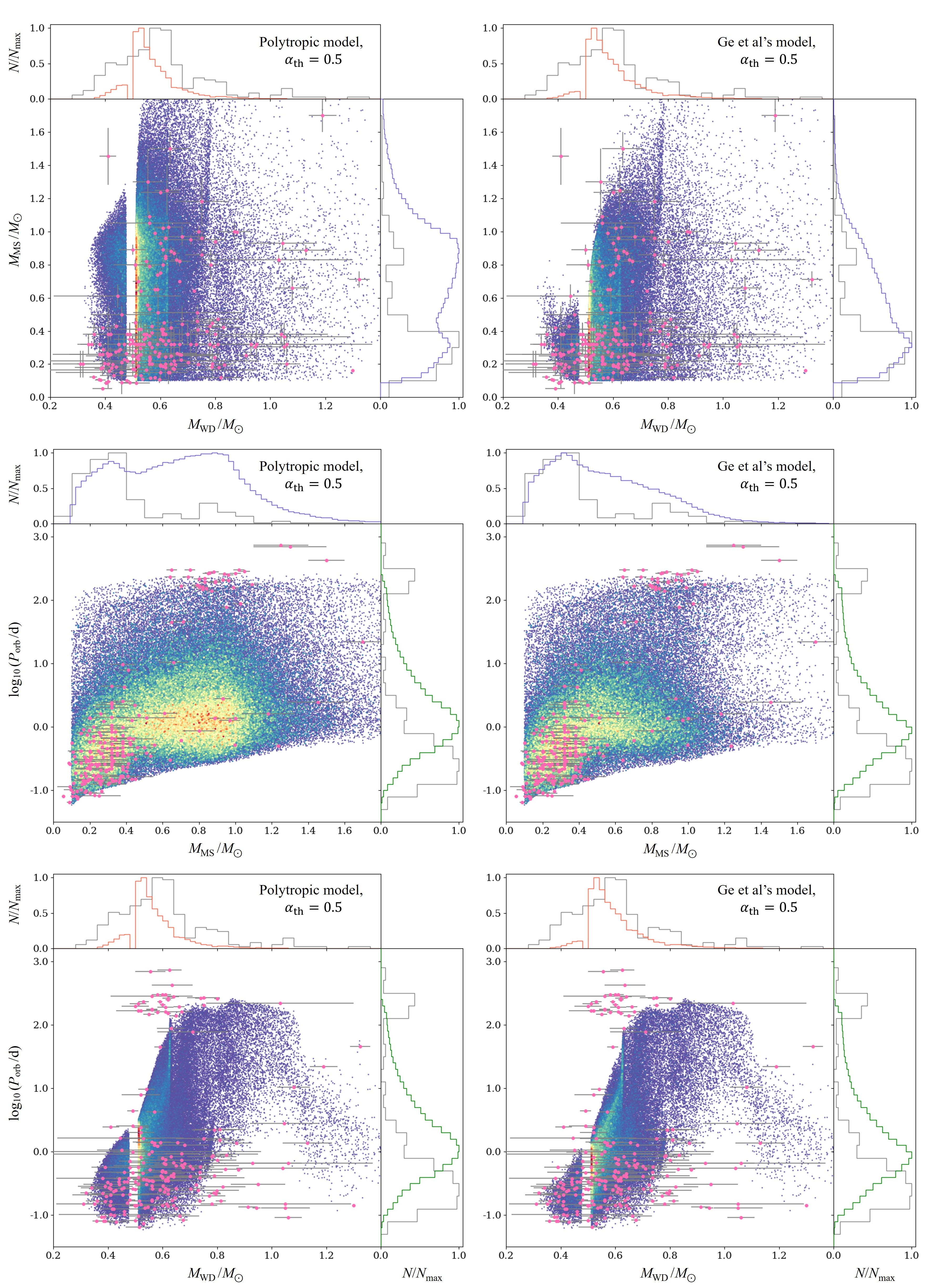}
\caption{The distribution of observed (pink dots, classic PCEBs, and long-period WD+MS binaries) and simulated PCEBs adopting half of the internal energy contribution, plotted in the $M_{\rm WD}-M_{\rm MS}$, $M_{\rm MS}-P_{\rm orb}$, and $M_{\rm WD}-P_{\rm orb}$ planes. The $\alpha_{\rm CE}$ is set to 0.25. The color scheme is identical to \autoref{fig:M_WD-M_MS}.\label{fig:alpha_th=0.5}}
\end{figure*}

We plot the results of our BPS simulations including internal energy contributions together with the observational sample in \autoref{fig:alpha_th=0.5}. We find that increasing $\alpha_{\rm th}$ has a similar effect to increasing $\alpha_{\rm CE}$: both facilitate envelope ejection, thereby broadening the WD mass distribution and increasing the total number of PCEBs. However, unlike $\alpha_{\rm CE}$, the inclusion of internal energy does not uniformly increase envelope ejection efficiency; instead, its effect depends strongly on the evolutionary state and initial mass of the star. For stars evolving to the late TP-AGB phase with initial masses of $3-5\,M_\odot$, ionization energy can dominate the total envelope binding energy, making the envelope binding energy very small (or even positive). This causes $\lambda$ to increase rapidly and eventually diverge to infinity \citep{2010ApJ...716..114X}. When MT initiates close to this regime, the envelope is easily ejected with minimal orbital shrinkage, forming wide-orbit PCEBs. However, this effect does not occur for stars with initial masses of $1-2\,M_\odot$, as their envelopes are more compact and have much lower values of $\lambda$. As a result, the long-period PCEBs formed in our models are exclusively systems with massive WDs, forming the long-period tail in the orbital period distribution shown in \autoref{fig:alpha_th=0.5}. Conversely, in models without internal energy, binaries still experience significant orbital shrinkage during CE evolution and therefore form shorter-period systems. Notably, even without any contribution from internal energy, models (a) and (d) can still form some massive WD systems with orbital periods of tens of days, such as IK Peg. This is consistent with previous work by \citet{2024A&A...686A..61B}.

Although the inclusion of internal energy in our models does not fully cover the parameter space of the observed long-period WD+MS systems, this does not imply that such systems cannot form via the CE channel; rather, it highlights a limitation of current BPS models. Detailed stellar evolution calculations show that MT during the TP-AGB phase typically occurs during thermal pulses, when the star expands rapidly and has a very low envelope binding energy \citep{2024A&A...687A..12B,2024PASP..136h4202Y}. Furthermore, additional studies have identified intense pulsation-driven envelope ejection operating in the late TP-AGB phase \citep{2026A&A...707A.342C}, which further lowers the energetic barrier to envelope ejection. Therefore, even a $1\,M_\odot$ star undergoing MT during a thermal pulse can form long-period PCEBs within a certain range of initial orbital separations. However, most BPS models do not resolve individual thermal pulses during the TP-AGB phase, instead using time-averaged stellar properties. Consequently, explaining the formation of long-period WD+MS binaries via CE evolution without accounting for thermal pulses on the TP-AGB is physically incomplete and unrealistic.

An important additional point is that, although we use a different value of $\alpha_{\rm th}$ here than in Section \ref{Sec:4}, the H. Ge et al. $q_{\rm c}$ prescription still produces a significant reduction in the number of PCEBs with solar-type MS companions compared to the polytropic $q_{\rm c}$ prescription. This demonstrates that the suppression of PCEBs with solar-type companions in our models is not dependent on a specific combination of $\alpha_{\rm th}$ and $\alpha_{\rm CE}$, but is a general result for all binaries undergoing CE evolution.

Furthermore, given that long-period WD+MS binaries typically have more massive companions and lie very close to the stable $M_{\rm WD}-P_{\rm orb}$ relation, another natural explanation for their origin is stable MT. A recent study by \citet{zheng2026massorbitalperioddistributionmassive} found that, at slightly low metallicity, intermediate-mass donors can form WD+MS binaries with orbital periods of hundreds of days via stable MT (\autoref{fig:obs fig}). The primary reason is that the progenitors of intermediate-mass stars exhibit distinct core properties during core growth: their cores remain nondegenerate prior to carbon--oxygen core formation, and thus, there is no well-defined core mass–radius relation \citep{zheng2026massorbitalperioddistributionmassive}. This mechanism provides an excellent explanation for the long-period WD+MS systems with massive companions in \autoref{fig:obs fig}, such as SLB1, SLB2, and SLB3 \citep{2018AJ....155..144K}.

Both CE evolution and stable MT can explain the formation of long-period WD+MS binaries to some extent, but each has its own problems. When we try to explain these systems using CE evolution with internal energy, we inevitably produce a large number of intermediate-period PCEBs with orbital periods of 10–100 days (see \autoref{fig:alpha_th=0.5}). However, almost no PCEBs are observed in this period range, and other observational evidence also supports the existence of this "period valley" \citep{2019MNRAS.484.5362A,2022MNRAS.512.2625L}. On the other hand, stable MT cannot avoid companion accretion and mass gain. This means that if we use stable MT to explain long-period WD+MS systems with low-mass companions ($\sim0.6\,M_\odot$), the mass ratio at the onset of MT would exceed 4, and it is unknown whether stable MT can still occur in this case. Furthermore, these long-period WD+MS systems also have higher eccentricities than expected, which contradicts the circular orbits predicted after stable MT \citep{2024PASP..136h4202Y}. 

Another proposed pathway is the grazing envelope evolution (GEE; \citealt{2015ApJ...800..114S,2018MNRAS.480.3195K,2018MNRAS.477.2584S}). In this scenario, the companion grazes the outer envelope of the giant and accretes material through a disk, launching jets that enhance envelope removal and may postpone or prevent the onset of a full CE phase. Thus, the GEE may produce postinteraction binaries with periods of several months to a few years \citep{2015ApJ...800..114S,2018MNRAS.480.3195K}. Periastron-enhanced mass loss during an eccentric GEE can also counteract tidal circularization and preserve a substantial eccentricity. Observational motivation comes from post-AGB binaries with similar orbital periods and jet-launching companions \citep{2024PASP..136h4202Y,2025A&A...698A.257I}. We do not model the GEE here, as no calibrated prescription currently relates the jet feedback to the final binary parameters, but it represents an additional possible channel for these long-period WD+MS systems.

This suggests that the observed wide-orbit WD+MS systems can originate from multiple formation channels, including CE evolution, stable MT, and possibly the GEE.

\subsection{Observational Selection Effects}\label{Sec:5.4}

Although we have attempted to compile a complete sample of PCEBs, significant observational selection effects remain. For classic short-period PCEBs detected via eclipses or spectroscopy, early-type (F/G/K) companions are much more luminous than the WD. The WD’s spectrum is therefore completely overwhelmed, and the binary is often misclassified as a single MS star in the Hertzsprung--Russell diagram. Additionally, the eclipse depth for these systems is too shallow to be detected in photometric surveys. For these reasons, PCEBs identified via eclipsing or spectroscopic surveys are strongly biased toward systems with M dwarf companions \citep{2011A&A...536A..43N,2012MNRAS.419..806R,2023MNRAS.521.1880B,2025A&A...695A.161S}. This may explain the tail of short-period systems with solar-type companions in our simulations, which is not seen in the observational sample in \autoref{fig:M_MS-P_orb}.

\begin{figure}[ht!]
\centering
\includegraphics[scale=0.65]{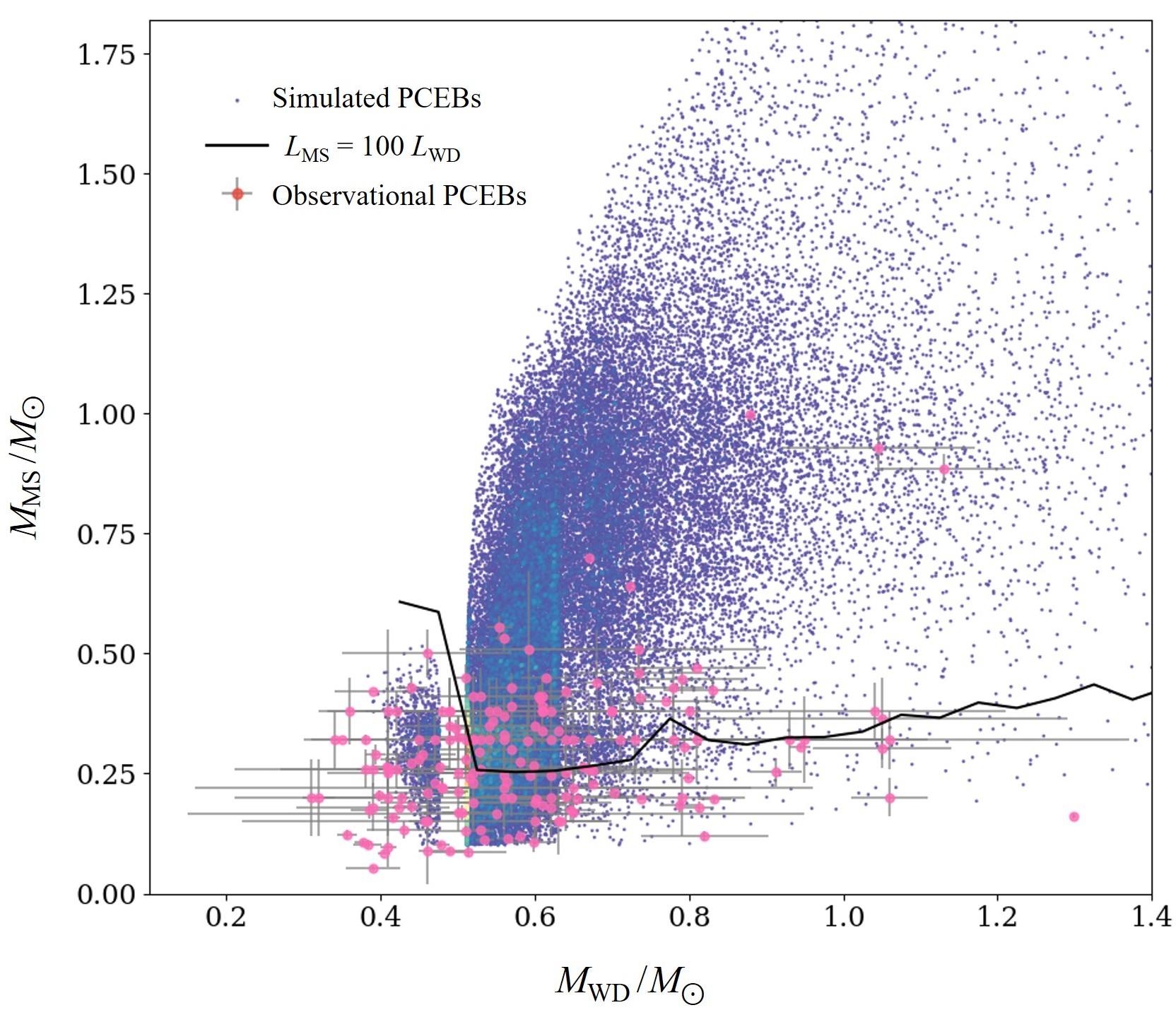}
\caption{The distribution of observed (pink dots) and simulated PCEBs in the $M_{\rm WD}-M_{\rm MS}$ plane. We exclude systems discovered via UV excess from this plot, as they are subject to totally different selection effects. The color scheme is identical to \autoref{fig:M_WD-M_MS}. The simulated PCEBs shown here are taken from model (a). The black solid line marks the locus where the MS companion luminosity is 100 times the mean WD luminosity in $0.05\,M_\odot$ WD mass bins, for WD masses spanning $0.35\,M_\odot$ to $1.4\,M_\odot$.\label{fig:selection eff}}
\end{figure}

To test whether observational selection effects, rather than our adopted $q_{\rm c}$ prescription, are the primary cause of the observed dearth of PCEBs with solar-type companions, we calculate the MS companion mass at which the companion luminosity equals 100 times the mean WD luminosity, binned in $0.05\,M_\odot$ intervals across a WD mass range of $0.35\,M_\odot$ to $1.4\,M_\odot$. We compare this derived threshold to the observed classic PCEB sample in \autoref{fig:selection eff}. However, it is important to note that individual WDs span a wide range of effective temperatures and therefore exhibit large luminosity scatter. As a result, the $L_{\rm MS}=100\,L_{\rm WD}$ threshold line only represents the average behavior in each mass bin, rather than a rigid cutoff for individual systems. The key takeaway from this figure lies in the contrast between C/O WDs and He WDs. C/O WDs have lower average luminosities, so their statistical threshold corresponds to lower companion masses. Even so, we observe many systems lying above this threshold, indicating that they should have been detectable. He WDs, by contrast, have significantly higher average luminosities, particularly hot young He WDs. For these systems, the $L_{\rm MS}=100\,L_{\rm WD}$ threshold corresponds to higher companion masses, meaning that even solar-type companions should not fully overwhelm the WD flux in optical surveys. If the polytropic $q_{\rm c}$ prescription indeed predicts a substantial population of He WD + FGK companion systems, at least a fraction of these systems should already have been detected in existing surveys.

In addition, although we attempt to exclude systems that may have experienced a second episode of MT, this is extremely challenging, particularly for systems within the CV period gap. This is because PCEBs and detached CVs in the period gap are inherently spectroscopically indistinguishable. According to \citet{2008MNRAS.389.1563D}, the number of detached CVs is far larger than the number of PCEBs (pre-CVs) in the period gap. This implies that our observational sample of PCEBs with orbital periods of 2–3 hr and M3.5–M4.5 companion spectral types may be significantly contaminated by detached CVs. Extreme caution is therefore warranted when comparing model predictions with observations in this specific parameter regime.

Overall, we conclude that observational selection effects alone are insufficient to explain the observed dearth of PCEBs with solar-type companions and that this explanation is not physically robust. Admittedly, this section presents a simplified statistical analysis and has some limitations in rigor. More complete, homogeneous observational samples are still required to place tighter, more accurate constraints on the underlying CE evolution physics.

\section{Conclusion and Summary}

Using BPS calculations with an updated mass-transfer stability criterion and a self-consistent envelope binding-energy prescription, we investigate the formation of PCEBs. Our main conclusions are as follows:

\begin{enumerate}
    \item The new mass-transfer stability criterion predicts significantly larger critical mass ratios for giant donors than traditional polytropic prescriptions. Its higher critical mass ratios for RGB/AGB donors prevent binaries with comparable masses from entering CE evolution, drastically reducing the predicted number of solar-type companion PCEBs.
    
    \item The model with $\alpha_{\rm CE}=0.25$ provides the best overall fit to the observed parameter distributions, consistent with previous studies showing most short-period PCEBs form via inefficient CE evolution.
    
    \item Neither stronger MB prescriptions nor observational selection effects can fully account for the observed deficit alone. A combination of enhanced MT stability, disrupted MB, and selection effects is required to reproduce the full PCEB population.
    
    \item A moderate contribution from envelope internal energy can produce some long-period WD+MS binaries, but cannot fully explain the observed wide-orbit population, suggesting that multiple formation channels may contribute.
\end{enumerate}

Overall, our study demonstrates that realistic mass-transfer stability criteria are essential for interpreting the observed PCEB population. These results place new constraints on CE evolution and have important implications for the formation of CVs, AM CVn stars, double WDs, and other compact binary populations.

\begin{acknowledgments}

This project is supported by the National Natural Science Foundation of China (NSFC Nos. 12525304, 12288102, 12125303, 12473033, 12333008, 12473034, 12422305), CAS Project for Young Scientists in Basic Research (YSBR-148), the Strategic Priority Research Program of the Chinese Academy of Sciences (grant Nos. XDB1160201, XDB1160303, XDB1160300, XDB1160000, XDB1160200), the National Key R\&D Program of China (No. 2021YFA1600403), Yunnan Revitalization Talent Support Program—Science \& Technology Champion Project (No. 202305AB350003), Yunnan Revitalization Talent Support Program—Young Talent Project, Yunnan Revitalization Talent Support Program "YunLing Scholar" project, Yunnan Fundamental Research Projects (Nos. 202401BC070007, 202401AT070139, 202601CJ070008), the MINECO grants PID2020-117252GB-I00, PID2023-148661NB-I00, and by the AGAUR/Generalitat de Catalunya grant SGR-386/2021, the PhD grant PRE2021-100503 funded by MICIU/AEI/10.13039/, the New Cornerstone Science Foundation through the XPLORER PRIZE, the International Centre of Supernovae (ICESUN) and Yunnan Key Laboratory of Supernova Research (Nos. 202302AN360001, 202505AV340004). We also thank the anonymous referee for the constructive and insightful comments, which have significantly improved the manuscript.

\end{acknowledgments}

\appendix

\section{Additional Observational Tables}

The observed physical parameters of our sampled PCEBs are summarized in \autoref{tab:observation}. Samples with corrected masses are listed in \autoref{tab:Corrected M_WD}.

\setcounter{table}{0}
\renewcommand{\thetable}{\Alph{section}\arabic{table}}       
\renewcommand{\theHtable}{\Alph{section}\arabic{table}}

\setcounter{figure}{0}
\renewcommand{\thefigure}{\Alph{section}\arabic{figure}}       
\renewcommand{\theHfigure}{\Alph{section}\arabic{figure}}

\startlongtable
\begin{deluxetable*}{lcccc}
\tabletypesize{\scriptsize}
\tablecaption{Observed properties of the PCEBs in our sample, sorted by orbital period.\label{tab:observation}}
\tablehead{
\colhead{Object} & \colhead{$M_{\rm WD}$} & \colhead{$M_{\rm MS}$} & \colhead{$P_{\rm orb}$} & \colhead{References} \\
\colhead{} & \colhead{($M_\odot$)} & \colhead{($M_\odot$)} & \colhead{({\rm day})} & \colhead{}
}
\startdata
SDSS   J085746.18+034255.3    & 0.514 ± 0.049 & 0.087 ± 0.012 & 0.065  & (11) \\
SDSS J0138-0016               & 0.530 ± 0.010 & 0.132 ± 0.003 & 0.073  & (11) \\
ZTF J163421.00-271321.7       & 0.430 ± 0.048 & 0.132 ± 0.018 & 0.078  & (3)  \\
ZTF J164441.18+243428.2       & 0.383 ± 0.019 & 0.103 ± 0.009 & 0.08   & (3)  \\
WD 0137-3457                  & 0.390 ± 0.035 & 0.053 ± 0.006 & 0.08   & (9)  \\
ZTF J041016.82-083419.5       & 0.357 ± 0.013 & 0.123 ± 0.009 & 0.081  & (3)  \\
SDSS 1611+4640                & 0.500 ± 0.056 & 0.250 ± 0.120 & 0.082  & (1)  \\
Gaia DR3 6765026180753366400  & 0.675 ± 0.016 & 0.227 ± 0.007 & 0.084  & (5)  \\
SDSS 0152-0058                & 0.560 ± 0.060 & 0.200 ± 0.080 & 0.09   & (1)  \\
SDSS 2112+1014                & 1.060 ± 0.050 & 0.200 ± 0.040 & 0.092  & (10) \\
ZTF J051902.06+092526.4       & 0.386 ± 0.024 & 0.174 ± 0.017 & 0.093  & (3)  \\
ZTF J102653.47-101330.3       & 0.377 ± 0.011 & 0.106 ± 0.007 & 0.093  & (3)  \\
Gaia DR3 4384149753578863744  & 0.662 ± 0.071 & 0.264 ± 0.006 & 0.096  & (5)  \\
ZTF J134151.70-062613.9       & 0.510 ± 0.037 & 0.129 ± 0.012 & 0.097  & (3)  \\
Gaia DR3 1088257649525365248  & 0.423 ± 0.008 & 0.178 ± 0.005 & 0.099  & (5)  \\
GD 448                        & 0.410 ± 0.010 & 0.096 ± 0.040 & 0.103  & (9)  \\
SDSS J2208+0037               & 0.460 ± 0.240 & 0.150 ± 0.032 & 0.103  & (11) \\
1355+0856                     & 0.460 ± 0.010 & 0.090 ± 0.070 & 0.114  & (11) \\
HS 2237+8154                  & 0.570 ± 0.100 & 0.300 ± 0.100 & 0.124  & (11) \\
SDSS J1210+3347               & 0.415 ± 0.010 & 0.158 ± 0.006 & 0.124  & (11) \\
SDSS 1435+3733                & 0.420 ± 0.050 & 0.260 ± 0.040 & 0.126  & (10) \\
M3-1                          & 0.65          & 0.170 ± 0.020 & 0.127  & (11) \\
Gaia DR3 739390405396929152   & 0.944 ± 0.014 & 0.305 ± 0.005 & 0.129  & (5)  \\
LAMOST J101356.33+272410.7    & 1.050 ± 0.090 & 0.303 ± 0.041 & 0.129  & (18) \\
NN Ser                        & 0.535 ± 0.012 & 0.111 ± 0.004 & 0.13   & (9)  \\
SDSS 0303-0054                & 0.912 ± 0.034 & 0.253 ± 0.029 & 0.134  & (9)  \\
Gaia DR3 155649614856576      & 0.546 ± 0.005 & 0.360 ± 0.005 & 0.135  & (5)  \\
CSS J090826.3+123648          & 0.62          & 0.18          & 0.139  & (11) \\
SDSS J1021+1744               & 0.500 ± 0.050 & 0.325 ± 0.045 & 0.14   & (11) \\
SDSS 0320-0638                & 0.640 ± 0.160 & 0.250 ± 0.120 & 0.141  & (1)  \\
ZTF J070458.08-020103.3       & 0.498 ± 0.014 & 0.343 ± 0.019 & 0.141  & (3)  \\
ZTF J094826.35+253810.6       & 0.505 ± 0.025 & 0.169 ± 0.015 & 0.142  & (3)  \\
SDSS J1151-0007               & 0.600 ± 0.100 & 0.190 ± 0.080 & 0.142  & (11) \\
ZTF J140702.57+211559.7       & 0.408 ± 0.016 & 0.265 ± 0.019 & 0.143  & (3)  \\
Gaia DR3 2273583445431091584  & 1.299 ± 0.002 & 0.161 ± 0.004 & 0.143  & (5)  \\
CSS J111647.8+294602          & 0.52          & 0.23          & 0.146  & (11) \\
LTT 560                       & 0.520 ± 0.120 & 0.190 ± 0.050 & 0.148  & (9)  \\
CSS 080502                    & 0.350 ± 0.040 & 0.32          & 0.149  & (9)  \\
SDSS 2123+0024                & 0.310 ± 0.100 & 0.200 ± 0.080 & 0.149  & (9)  \\
ARSco                         & 1.050 ± 0.240 & 0.365 ± 0.085 & 0.149  & (11) \\
EC 13471-1258                 & 0.780 ± 0.040 & 0.430 ± 0.040 & 0.15   & (15) \\
Gaia DR3 3612227169936143360  & 0.614 ± 0.032 & 0.450 ± 0.006 & 0.151  & (5)  \\
CSS J081158.6+311959          & 0.47          & 0.23          & 0.156  & (11) \\
SDSS 1529+0020                & 0.390 ± 0.020 & 0.260 ± 0.040 & 0.165  & (10) \\
PG 1458+172                   & 0.41          & 0.380 ± 0.170 & 0.165  & (11) \\
ZTF J130228.34-003200.2       & 0.813 ± 0.019 & 0.180 ± 0.011 & 0.166  & (3)  \\
SDSS 1411+1028                & 0.360 ± 0.040 & 0.380 ± 0.070 & 0.167  & (11) \\
ZTF J065103.70+145246.2       & 0.515 ± 0.020 & 0.242 ± 0.019 & 0.168  & (3)  \\
ZTF J140423.86+065557.7       & 0.736 ± 0.016 & 0.409 ± 0.023 & 0.168  & (3)  \\
TMTS J15530469+4457458        & 0.560 ± 0.090 & 0.370 ± 0.020 & 0.168  & (13) \\
ZTF J064242.41+131427.6       & 0.634 ± 0.010 & 0.150 ± 0.008 & 0.171  & (3)  \\
MS Peg                        & 0.480 ± 0.020 & 0.220 ± 0.020 & 0.174  & (9)  \\
NGC 6337                      & 0.56          & 0.245 ± 0.105 & 0.174  & (11) \\
SDSS 1548+4057                & 0.646 ± 0.032 & 0.174 ± 0.027 & 0.185  & (9)  \\
CSS J090119.2+114254          & 0.58          & 0.12          & 0.187  & (11) \\
ZTF J080542.98-143036.3       & 0.393 ± 0.013 & 0.290 ± 0.021 & 0.198  & (3)  \\
BPM 71214                     & 0.770 ± 0.060 & 0.4           & 0.202  & (11) \\
Gaia DR3 2562060180904793344  & 0.454 ± 0.043 & 0.289 ± 0.023 & 0.21   & (5)  \\
SDSS 2216+0102                & 0.410 ± 0.140 & 0.260 ± 0.040 & 0.21   & (10) \\
SDSS 0238-0005                & 0.480 ± 0.150 & 0.380 ± 0.010 & 0.212  & (10) \\
ZTF J071843.68-085232.1       & 0.794 ± 0.019 & 0.306 ± 0.020 & 0.216  & (3)  \\
SDSS 1625+6400                & 0.620 ± 0.100 & 0.200 ± 0.080 & 0.218  & (1)  \\
SDSS 2132+0031                & 0.380 ± 0.040 & 0.320 ± 0.010 & 0.222  & (9)  \\
SDSS 1844+4120                & 0.320 ± 0.020 & 0.200 ± 0.080 & 0.226  & (1)  \\
ZTF J052848.24+215629.0       & 0.787 ± 0.025 & 0.184 ± 0.014 & 0.226  & (3)  \\
ZTF J102254.00-080327.3       & 0.606 ± 0.026 & 0.405 ± 0.030 & 0.231  & (3)  \\
SDSS J1028+0931               & 0.420 ± 0.040 & 0.380 ± 0.040 & 0.235  & (11) \\
Gaia DR3 1177642000628467072  & 0.799 ± 0.006 & 0.240 ± 0.003 & 0.239  & (5)  \\
SDSS 1231-0310                & 0.730 ± 0.150 & 0.320 ± 0.090 & 0.244  & (1)  \\
ZTF J104906.96-175530.7       & 0.427 ± 0.009 & 0.199 ± 0.011 & 0.245  & (3)  \\
Gaia DR3 1245093225062192640  & 0.620 ± 0.020 & 0.245 ± 0.004 & 0.248  & (5)  \\
SDSS 1348+1834                & 0.590 ± 0.040 & 0.319 ± 0.060 & 0.249  & (9)  \\
ZTF J162644.18-101854.3       & 0.500 ± 0.014 & 0.213 ± 0.012 & 0.253  & (3)  \\
ZTF J063954.70+191958.0       & 0.702 ± 0.010 & 0.210 ± 0.011 & 0.259  & (3)  \\
LM Com                        & 0.450 ± 0.050 & 0.280 ± 0.050 & 0.259  & (9)  \\
ZTF J140036.65+081447.4       & 0.563 ± 0.009 & 0.232 ± 0.012 & 0.26   & (3)  \\
SDSS 103736.57+013905.11      & 0.490 ± 0.020 & 0.089 ± 0.008 & 0.26   & (11) \\
SDSS 2240-0935                & 0.410 ± 0.080 & 0.250 ± 0.120 & 0.261  & (9)  \\
SDSS J0314-0111               & 0.650 ± 0.100 & 0.320 ± 0.090 & 0.263  & (11) \\
HS 1857+5144                  & 0.610 ± 0.040 & 0.410 ± 0.030 & 0.266  & (11) \\
SDSS 1731+6233                & 0.340 ± 0.040 & 0.320 ± 0.060 & 0.268  & (10) \\
ZTF J180256.45-005458.3       & 0.457 ± 0.020 & 0.150 ± 0.011 & 0.269  & (3)  \\
ZTF J140537.34+103919.0       & 0.404 ± 0.008 & 0.085 ± 0.005 & 0.271  & (3)  \\
SDSS 1006+0044                & 0.820 ± 0.083 & 0.120 ± 0.010 & 0.28   & (1)  \\
CC Cet                        & 0.390 ± 0.100 & 0.180 ± 0.050 & 0.284  & (9)  \\
SDSS J1136+0409               & 0.601 ± 0.036 & 0.196 ± 0.085 & 0.287  & (11) \\
SDSS J212531-010745           & 0.560 ± 0.060 & 0.330 ± 0.080 & 0.29   & (11) \\
GPX-TF16E-48                  & 0.723         & 0.64          & 0.298  & (17) \\
RR Cae                        & 0.440 ± 0.022 & 0.182 ± 0.013 & 0.303  & (9)  \\
SDSS 1611+0103                & 0.410 ± 0.100 & 0.200 ± 0.080 & 0.304  & (1)  \\
SDSS 0833+0702                & 0.540 ± 0.070 & 0.320 ± 0.060 & 0.304  & (10) \\
SDSS 1300+1908                & 0.950 ± 0.103 & 0.320 ± 0.090 & 0.308  & (1)  \\
ZTF J195456.71+101937.5       & 0.510 ± 0.014 & 0.450 ± 0.027 & 0.31   & (3)  \\
ZTF J053708.26-245014.6       & 0.398 ± 0.008 & 0.204 ± 0.012 & 0.328  & (3)  \\
SDSS 0110+1326                & 0.470 ± 0.020 & 0.320 ± 0.050 & 0.333  & (10) \\
SDSS 1724+5620                & 0.460 ± 0.050 & 0.210 ± 0.030 & 0.333  & (11) \\
SDSS J1212-0123               & 0.439 ± 0.002 & 0.273 ± 0.002 & 0.336  & (11) \\
BPM 6502                      & 0.500 ± 0.050 & 0.170 ± 0.010 & 0.337  & (9)  \\
GK Vir                        & 0.564 ± 0.014 & 0.116 ± 0.003 & 0.344  & (11) \\
SDSS 1105+3851                & 0.550 ± 0.040 & 0.380 ± 0.070 & 0.345  & (1)  \\
ZTF J061530.96+051041.8       & 0.560 ± 0.011 & 0.533 ± 0.030 & 0.348  & (3)  \\
EC 14329-1625                 & 0.620 ± 0.110 & 0.380 ± 0.070 & 0.35   & (9)  \\
KIC 10544976                  & 0.610 ± 0.040 & 0.390 ± 0.030 & 0.35   & (11) \\
PTFEB 11.441                  & 0.540 ± 0.050 & 0.350 ± 0.050 & 0.359  & (11) \\
EC 12477-1738                 & 0.610 ± 0.080 & 0.380 ± 0.070 & 0.362  & (9)  \\
DE CVn                        & 0.530 ± 0.040 & 0.410 ± 0.060 & 0.364  & (10) \\
SDSS J0848+2320               & 0.440 ± 0.020 & 0.430 ± 0.022 & 0.372  & (11) \\
Gaia DR3 4333046892662353152  & 0.476 ± 0.013 & 0.264 ± 0.008 & 0.381  & (5)  \\
SDSS 1047+0523                & 0.380 ± 0.170 & 0.260 ± 0.040 & 0.382  & (10) \\
SDSS 1143+0009                & 0.620 ± 0.070 & 0.320 ± 0.010 & 0.386  & (9)  \\
PTFEB 28.235                  & 0.600 ± 0.060 & 0.350 ± 0.050 & 0.386  & (11) \\
LAMOST J035916 + 400732       & 0.735 ± 0.165 & 0.510 ± 0.050 & 0.39   & (16) \\
SDSS 0949+0322                & 0.510 ± 0.080 & 0.320 ± 0.060 & 0.396  & (10) \\
SDSS J1316-0037               & 0.640 ± 0.110 & 0.202 ± 0.010 & 0.403  & (11) \\
TIC 60040774                  & 0.598 ± 0.029 & 0.107 ± 0.020 & 0.405  & (14) \\
SDSS 2114-0103                & 0.700 ± 0.070 & 0.380 ± 0.010 & 0.411  & (10) \\
SDSS 1523+4604                & 0.630 ± 0.064 & 0.150 ± 0.070 & 0.414  & (1)  \\
SDSS 1608+0851                & 0.790 ± 0.083 & 0.200 ± 0.080 & 0.414  & (1)  \\
Gaia DR3 3070875851133903232  & 0.593 ± 0.012 & 0.247 ± 0.005 & 0.419  & (5)  \\
SDSS 2120-0058                & 0.640 ± 0.040 & 0.320 ± 0.060 & 0.449  & (10) \\
ZTF J071759.04+113630.2       & 0.528 ± 0.017 & 0.295 ± 0.021 & 0.453  & (3)  \\
PTFEB 28.852                  & 0.490 ± 0.060 & 0.350 ± 0.050 & 0.462  & (11) \\
EC 13349-3237                 & 0.460 ± 0.110 & 0.500 ± 0.050 & 0.47   & (9)  \\
SDSS 0807+0724                & 0.610 ± 0.100 & 0.380 ± 0.070 & 0.477  & (1)  \\
LAMOST J035540 + 381550       & 0.735 ± 0.085 & 0.460 ± 0.040 & 0.477  & (16) \\
Gaia DR3   627874465774248704 & 0.598 ± 0.011 & 0.266 ± 0.004 & 0.481  & (5)  \\
TYC 6760-497-1                & 0.600 ± 0.070 & 1.235 ± 0.015 & 0.499  & (11) \\
RX J2130.6+4710               & 0.554 ± 0.017 & 0.555 ± 0.023 & 0.521  & (9)  \\
V471 Tau                      & 0.878 ± 0.001 & 0.998 ± 0.002 & 0.521  & (11) \\
2M 10243847+1624582           & 0.830 ± 0.063 & 0.423 ± 0.027 & 0.526  & (11) \\
SDSS 0301+0502                & 0.710 ± 0.064 & 0.320 ± 0.090 & 0.539  & (1)  \\
SDSS 1429+5759                & 1.040 ± 0.170 & 0.380 ± 0.060 & 0.545  & (9)  \\
ZTF J125620.57+211725.8       & 0.479 ± 0.010 & 0.101 ± 0.005 & 0.556  & (3)  \\
HZ 9                          & 0.510 ± 0.100 & 0.280 ± 0.040 & 0.564  & (9)  \\
UX CVn                        & 0.390 ± 0.050 & 0.42          & 0.57   & (9)  \\
Gaia DR3 3639624796381535488  & 0.675 ± 0.115 & 0.256 ± 0.006 & 0.571  & (5)  \\
Gaia DR3 6916925365694417152  & 0.738 ± 0.005 & 0.198 ± 0.002 & 0.578  & (5)  \\
SDSS 1524+5040                & 0.700 ± 0.040 & 0.380 ± 0.070 & 0.59   & (11) \\
UZ Sex                        & 0.650 ± 0.230 & 0.220 ± 0.050 & 0.597  & (9)  \\
Hen 2-11                      & 0.67          & 0.7           & 0.609  & (11) \\
TIC 460388167                 & 0.610 ± 0.040 & 0.340 ± 0.010 & 0.636  & (4)  \\
SDSS 2149-0717                & 0.680 ± 0.020 & 0.440 ± 0.110 & 0.644  & (1)  \\
SDSS 2339-0020                & 0.930 ± 0.180 & 0.320 ± 0.060 & 0.655  & (10) \\
ZTF J063808.71+091027.4       & 0.605 ± 0.012 & 0.411 ± 0.023 & 0.658  & (3)  \\
SDSS 1558+2642                & 1.060 ± 0.310 & 0.320 ± 0.060 & 0.661  & (10) \\
EG Uma                        & 0.640 ± 0.030 & 0.420 ± 0.040 & 0.668  & (9)  \\
SDSS 1718+6101                & 0.530 ± 0.060 & 0.320 ± 0.060 & 0.673  & (10) \\
RE J2013+4002                 & 0.480 ± 0.040 & 0.220 ± 0.020 & 0.706  & (11) \\
SDSS 0246+0041                & 0.800 ± 0.070 & 0.380 ± 0.010 & 0.728  & (10) \\
SDSS 1414-0132                & 0.670 ± 0.150 & 0.260 ± 0.040 & 0.728  & (10) \\
WD 2009+622                   & 0.610 ± 0.030 & 0.185 ± 0.001 & 0.741  & (11) \\
RE J1016-0520                 & 0.600 ± 0.020 & 0.150 ± 0.020 & 0.789  & (9)  \\
SDSS 1705+2109                & 0.520 ± 0.050 & 0.250 ± 0.120 & 0.815  & (9)  \\
HS 1136+6646                  & 0.630 ± 0.050 & 0.34          & 0.836  & (11) \\
SDSS 1437+5737                & 0.780 ± 0.090 & 0.320 ± 0.090 & 0.839  & (1)  \\
TYC 110-755-1                 & 0.784 ± 0.030 & 0.800 ± 0.090 & 0.858  & (7)  \\
Gaia DR3 2155188926705745536  & 0.833 ± 0.021 & 0.196 ± 0.006 & 0.883  & (5)  \\
SDSS J0225+0054               & 0.550 ± 0.400 & 0.166 ± 0.015 & 0.921  & (11) \\
SDSS 1313+0237                & 0.670 ± 0.060 & 0.320 ± 0.090 & 0.93   & (1)  \\
SDSS 2045-0509                & 0.490 ± 0.070 & 0.380 ± 0.070 & 0.98   & (1)  \\
Abell65                       & 0.560 ± 0.400 & 0.220 ± 0.040 & 1.004  & (11) \\
SDSS 1506-0120                & 0.450 ± 0.090 & 0.320 ± 0.060 & 1.051  & (10) \\
ZTF J122009.98+082155.0       & 0.580 ± 0.018 & 0.275 ± 0.020 & 1.233  & (3)  \\
IN CMa                        & 0.570 ± 0.030 & 0.430 ± 0.030 & 1.26   & (9)  \\
TYC 4962-1205-1               & 0.680 ± 0.090 & 0.969 ± 0.058 & 1.28   & (11) \\
SDSS 1519+3536                & 0.570 ± 0.030 & 0.200 ± 0.040 & 1.367  & (10) \\
CPD-65 264                    & 0.870 ± 0.060 & 1.000 ± 0.050 & 1.37   & (8)  \\
J0144+5106                    & 1.145 ± 0.095 & 0.894 ± 0.023 & 1.371  & (19) \\
KOI-256                       & 0.592 ± 0.089 & 0.510 ± 0.160 & 1.379  & (11) \\
Gaia DR3 1128036811987813888  & 0.654 ± 0.008 & 0.197 ± 0.003 & 1.382  & (5)  \\
SDSS 1528+3443                & 0.560 ± 0.070 & 0.320 ± 0.090 & 1.411  & (1)  \\
SDSS 1439-0106                & 0.810 ± 0.090 & 0.47          & 1.523  & (1)  \\
SDSS 1646+4223                & 0.530 ± 0.060 & 0.260 ± 0.040 & 1.595  & (10) \\
TYC 1380-957-1                & 0.750 ± 0.100 & 1.181 ± 0.145 & 1.613  & (11) \\
TYC 3858-1215-1               & 0.445 ± 0.231 & 0.610 ± 0.070 & 1.642  & (7)  \\
SDSS 0305-0749                & 0.520 ± 0.050 & 0.320 ± 0.090 & 2.019  & (1)  \\
2M 10552625+4729228           & 0.790 ± 0.081 & 0.446 ± 0.029 & 2.187  & (11) \\
SDSS 1623+6306                & 0.810 ± 0.090 & 0.320 ± 0.090 & 2.232  & (1)  \\
SDSS 0924+0024                & 0.520 ± 0.030 & 0.320 ± 0.060 & 2.404  & (10) \\
TYC 4700-815-1                & 0.410 ± 0.030 & 1.454 ± 0.171 & 2.467  & (11) \\
SDSS 2318-0935                & 0.490 ± 0.060 & 0.380 ± 0.070 & 2.534  & (9)  \\
J2013+1734                    & 1.075 ± 0.125 & 0.975 ± 0.051 & 2.757  & (19) \\
Feige 24                      & 0.570 ± 0.030 & 0.390 ± 0.020 & 4.232  & (9)  \\
SDSS 1434+5335                & 0.490 ± 0.020 & 0.320 ± 0.060 & 4.357  & (10) \\
SDSS J1211-0249               & 0.520 ± 0.070 & 0.410 ± 0.050 & 7.818  & (2)  \\
SDSS J2221+0029               & 0.540 ± 0.030 & 0.380 ± 0.070 & 9.588  & (2)  \\
2M 07515777+1807352           & 1.080 ± 0.060 & 0.660 ± 0.080 & 10.299 & (6)  \\
IK Peg                        & 1.190 ± 0.050 & 1.700 ± 0.100 & 21.722 & (9)  \\
J1314+3818                    & 1.324 ± 0.037 & 0.712 ± 0.049 & 45.519 & (12) \\
\enddata
\end{deluxetable*}
{\footnotesize
\noindent
\textbf{References---}
(1) \citet{2011A&A...536A..43N}; (2) \citet{2012MNRAS.423..320R}; (3) \citet{2023MNRAS.521.1880B}; (4) \citet{2026ApJ..1003..145B}; (5) \citet{2026PASP..138c4202S}; (6) \citet{2026AJ....171..159M}; (7) \citet{2022MNRAS.512.1843H}; (8) \citet{2022MNRAS.517.2867H}; (9) \citet{2010A&A...520A..86Z}; (10) \citet{2011A&A...536A..42Z}; (11) \citet{2021ApJ...920...86K}; (12) \citet{2024MNRAS.52711719Y}; (13) \citet{2025A&A...698A..81L}; (14) \citet{2022MNRAS.516.1183P}; (15) \citet{2009MNRAS.394..978P}; (16) \citet{2024MNRAS.532.1718Q}; (17) \citet{2020MNRAS.493.5208K}; (18) \citet{2025ApJ...984...42H}; (19) \citet{2026PASJ...78..382S}} 

\begin{deluxetable}{ccccccc}[ht!]
\tabletypesize{\scriptsize}
\tablecaption{Corrected white dwarf masses for systems affected by the log g upturn problem.\label{tab:Corrected M_WD}}
\tablehead{
\colhead{Object} & \colhead{$T_{\rm eff}$} & \colhead{${\rm log\,g}$} & \colhead{$M_{\rm WD}$} & \colhead{New $T_{\rm eff}$} & \colhead{New Log\,g} & \colhead{New $M_{\rm WD}$} \\
\colhead{} & \colhead{({\rm K})} & \colhead{} & \colhead{($M_\odot$)} & \colhead{({\rm K})} & \colhead{} & \colhead{($M_\odot$)}
}
\startdata
SDSS 0152–0058 & 8773  & 8.19 & 0.72  & 8775  & 7.94 & 0.56 \\
SDSS 0301+0502 & 11,109 & 8.43 & 0.875 & 11,162 & 8.19 & 0.71 \\
SDSS 0320–0638 & 11,173 & 8.3  & 0.79  & 11,264 & 8.08 & 0.64 \\
SDSS 1006+0044 & 7819  & 8.52 & 0.93  & 7823  & 8.38 & 0.82 \\
SDSS 1105+3851 & 10,548 & 8.17 & 0.71  & 10,520 & 7.9  & 0.55 \\
SDSS 1231–0310 & 10,073 & 8.51 & 0.93  & 10,022 & 8.23 & 0.73 \\
SDSS 1300+1908 & 8673  & 8.81 & 1.09  & 8657  & 8.57 & 0.95 \\
SDSS 1523+4604 & 8378  & 8.28 & 0.78  & 8396  & 8.07 & 0.63 \\
SDSS 1608+0851 & 9844  & 8.6  & 0.98  & 9794  & 8.32 & 0.79 \\
SDSS 1611+0103 & 10,189 & 7.81 & 0.49  & 10,159 & 7.55 & 0.41 \\
SDSS 1611+4640 & 10,307 & 8.04 & 0.63  & 10,268 & 7.78 & 0.5  \\
SDSS 1623+6306 & 9731  & 8.64 & 1     & 9682  & 8.36 & 0.81 \\
SDSS 1625+6400 & 8773  & 8.3  & 0.79  & 8779  & 8.05 & 0.62 \\
SDSS 1844+4120 & 7554  & 7.49 & 0.34  & 7575  & 7.33 & 0.32
\enddata
\end{deluxetable}

\section{Evolution Pathways of Typical systems}
\setcounter{table}{0}

We list the full evolutionary track of a typical system by adopting the default critical mass ratio in \autoref{tab:default_qc} and the critical mass ratio from H. Ge et al.'s model in \autoref{tab:ge_qc}.

\begin{deluxetable}{cccccccc}[ht!]
\tabletypesize{\scriptsize}
\tablecaption{Full evolutionary track of a representative binary system from our simulation, adopting the default $q_{\rm c}$ prescription of BSE \citep{2002MNRAS.329..897H}.\label{tab:default_qc}}
\tablehead{
\colhead{Time} & \colhead{$M_1$} & \colhead{$M_2$} & \colhead{${\rm Type}_1$} & \colhead{${\rm Type}_2$} & \colhead{$P_{\rm orb}$} & \colhead{$R_1/R_{\rm L1}$} & \colhead{Event Type} \\
\colhead{(Myr)} & \colhead{($M_\odot$)} & \colhead{($M_\odot$)} & \colhead{} & \colhead{} & \colhead{({\rm days})} & \colhead{} & \colhead{}
}
\startdata
0        & 1.5568 & 1.058  & MS    & MS & 434.64 & 0.0107 & ZAMS                \\
2438.334 & 1.5568 & 1.058  & HG    & MS & 434.64 & 0.0205 & Stellar type change \\
2493.855 & 1.5566 & 1.058  & RGB   & MS & 434.64 & 0.0265 & Stellar type change \\
2622.249 & 1.515  & 1.0609 & RGB   & MS & 398.11 & 1.0005 & RLOF ($q_{\rm c}=0.83$)      \\
2622.249 & 0.4549 & 1.0609 & He WD & MS & 0.72   & 1.0005 & CE evolution        \\
2622.249 & 0.4549 & 1.0609 & He WD & MS & 0.72   & 0.0125 & End RLOF            \\
3245.026 & 0.4549 & 1.0609 & He WD & MS & 0.7    & 0.0128 & Reach $t_{\rm max}$
\enddata
\tablecomments{Hertzsprung gap (HG); Roche lobe overflow (RLOF).}
\end{deluxetable}

\begin{deluxetable}{cccccccc}[ht!]
\tabletypesize{\scriptsize}
\tablecaption{Full evolutionary track of a binary system with identical initial parameters to \autoref{tab:default_qc}, adopting the $q_{\rm c}$ prescription from \citet{2023ApJ...945....7G}.\label{tab:ge_qc}}
\tablehead{
\colhead{Time} & \colhead{$M_1$} & \colhead{$M_2$} & \colhead{${\rm Type}_1$} & \colhead{${\rm Type}_2$} & \colhead{$P_{\rm orb}$} & \colhead{$R_1/R_{\rm L1}$} & \colhead{Event Type} \\
\colhead{(Myr)} & \colhead{($M_\odot$)} & \colhead{($M_\odot$)} & \colhead{} & \colhead{} & \colhead{({\rm days})} & \colhead{} & \colhead{}
}
\startdata
0        & 1.5568 & 1.058  & MS     & MS & 434.64 & 0.0107 & ZAMS                       \\
2438.334 & 1.5568 & 1.058  & HG     & MS & 434.64 & 0.0205 & Stellar type change        \\
2493.855 & 1.5566 & 1.058  & RGB    & MS & 434.64 & 0.0265 & Stellar type change        \\
2622.249 & 1.515  & 1.0609 & RGB    & MS & 398.11 & 1.0005 & RLOF ($q_{\rm c}=2$)       \\
2622.484 & 1.5036 & 1.0642 & CHeB   & MS & 398.11 & 0.097  & Stellar type change        \\
2622.484 & 1.5036 & 1.0642 & CHeB   & MS & 398.11 & 0.097  & End RLOF                   \\
2754.083 & 1.4771 & 1.0645 & E-AGB  & MS & 405.42 & 0.1898 & Stellar type change        \\
2757.601 & 1.4615 & 1.0658 & E-AGB  & MS & 379.85 & 1.0013 & RLOF ($q_{\rm c}=2$)       \\
2757.883 & 1.3725 & 1.1431 & TP-AGB & MS & 362.05 & 1.8268 & Stellar type change        \\
2758.116 & 1.0064 & 1.4952 & TP-AGB & MS & 409.07 & 2.1639 & Blue straggler   formation \\
2758.672 & 0.5612 & 1.9007 & TP-AGB & MS & 781.61 & 0.8535 & End RLOF                   \\
2758.789 & 0.5568 & 1.9022 & C/O WD & MS & 785.27 & 0.0001 & Stellar type change        \\
3245.026 & 0.5568 & 1.9022 & C/O WD & MS & 785.27 & 0.0001 & Reach $t_{\rm max}$
\enddata
\tablecomments{Core helium burning (CHeB); early asymptotic giant branch (E-AGB).}
\end{deluxetable}

\section{Comparison with different MB prescriptions}

\begin{figure*}[ht!]
\centering
\includegraphics[scale=0.9]{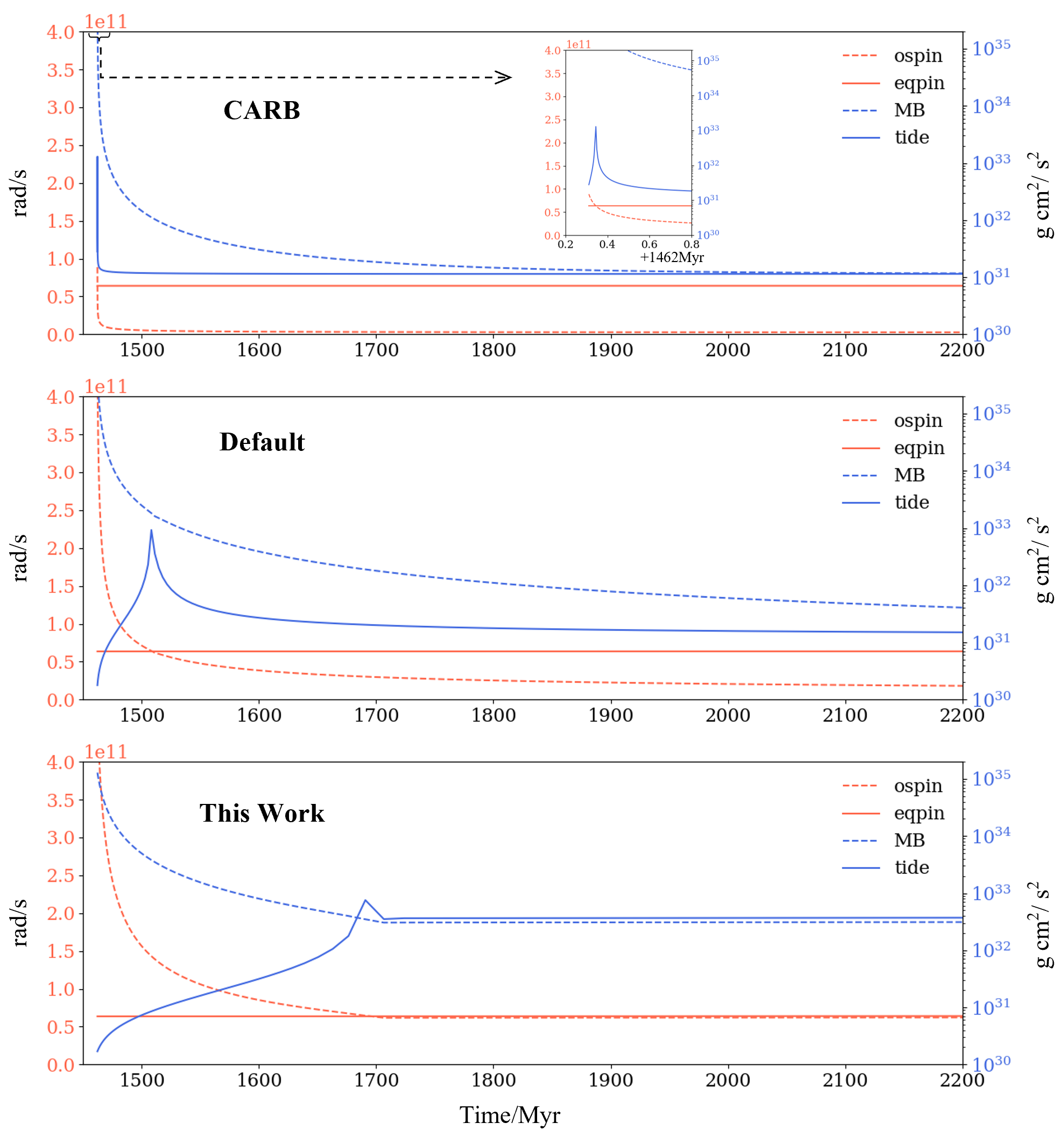}
\caption{The angular momentum evolution and spin angular velocity as a function of time for a representative PCEB with $M_{\rm WD}=0.5747\,M_\odot$, $M_{\rm MS}=0.3666\,M_\odot$, and a post-CE orbital period of $\sim$1.14 days. Blue-dashed and solid lines show the angular momentum change rate from MB and tides, respectively. Red-dashed and solid lines show the instantaneous spin angular velocity and the tidal synchronization angular velocity, respectively.\label{fig:MB compare}}
\end{figure*}

\bibliography{sample701}{}

@ARTICLE{2024PrPNP.13404083C,
       author = {{Chen}, Xuefei and {Liu}, Zhengwei and {Han}, Zhanwen},
        title = "{Binary stars in the new millennium}",
      journal = {Progress in Particle and Nuclear Physics},
         year = 2024,
        month = jan,
       volume = {134},
          eid = {104083},
        pages = {104083},
          doi = {10.1016/j.ppnp.2023.104083},
archivePrefix = {arXiv},
       eprint = {2311.11454},
 primaryClass = {astro-ph.SR},
       adsurl = {https://ui.adsabs.harvard.edu/abs/2024PrPNP.13404083C}
}

@ARTICLE{2010A&A...520A..86Z,
       author = {{Zorotovic}, M. and {Schreiber}, M.~R. and {G{\"a}nsicke}, B.~T. and {Nebot G{\'o}mez-Mor{\'a}n}, A.},
        title = "{Post-common-envelope binaries from SDSS. IX: Constraining the common-envelope efficiency}",
      journal = {\aap},
         year = 2010,
        month = sep,
       volume = {520},
          eid = {A86},
        pages = {A86},
          doi = {10.1051/0004-6361/200913658},
archivePrefix = {arXiv},
       eprint = {1006.1621},
 primaryClass = {astro-ph.SR},
       adsurl = {https://ui.adsabs.harvard.edu/abs/2010A&A...520A..86Z}
}

@ARTICLE{2011A&A...536A..42Z,
       author = {{Zorotovic}, M. and {Schreiber}, M.~R. and {G{\"a}nsicke}, B.~T.},
        title = "{Post common envelope binaries from SDSS. XI. The white dwarf mass distributions of CVs and pre-CVs}",
      journal = {\aap},
         year = 2011,
        month = dec,
       volume = {536},
          eid = {A42},
        pages = {A42},
          doi = {10.1051/0004-6361/201116626},
archivePrefix = {arXiv},
       eprint = {1108.4600},
 primaryClass = {astro-ph.SR},
       adsurl = {https://ui.adsabs.harvard.edu/abs/2011A&A...536A..42Z}
}

@ARTICLE{2011A&A...536A..43N,
       author = {{Nebot G{\'o}mez-Mor{\'a}n}, A. and {G{\"a}nsicke}, B.~T. and {Schreiber}, M.~R. and {Rebassa-Mansergas}, A. and {Schwope}, A.~D. and {Southworth}, J. and {Aungwerojwit}, A. and {Bothe}, M. and {Davis}, P.~J. and {Kolb}, U. and {M{\"u}ller}, M. and {Papadaki}, C. and {Pyrzas}, S. and {Rabitz}, A. and {Rodr{\'\i}guez-Gil}, P. and {Schmidtobreick}, L. and {Schwarz}, R. and {Tappert}, C. and {Toloza}, O. and {Vogel}, J. and {Zorotovic}, M.},
        title = "{Post common envelope binaries from SDSS. XII. The orbital period distribution}",
      journal = {\aap},
         year = 2011,
        month = dec,
       volume = {536},
          eid = {A43},
        pages = {A43},
          doi = {10.1051/0004-6361/201117514},
archivePrefix = {arXiv},
       eprint = {1109.6662},
 primaryClass = {astro-ph.SR},
       adsurl = {https://ui.adsabs.harvard.edu/abs/2011A&A...536A..43N}
}

@ARTICLE{2010A&A...513L...7S,
       author = {{Schreiber}, M.~R. and {G{\"a}nsicke}, B.~T. and {Rebassa-Mansergas}, A. and {Nebot Gomez-Moran}, A. and {Southworth}, J. and {Schwope}, A.~D. and {M{\"u}ller}, M. and {Papadaki}, C. and {Pyrzas}, S. and {Rabitz}, A. and {Rodr{\'\i}guez-Gil}, P. and {Schmidtobreick}, L. and {Schwarz}, R. and {Tappert}, C. and {Toloza}, O. and {Vogel}, J. and {Zorotovic}, M.},
        title = "{Post common envelope binaries from SDSS. VIII. Evidence for disrupted magnetic braking}",
      journal = {\aap},
         year = 2010,
        month = apr,
       volume = {513},
          eid = {L7},
        pages = {L7},
          doi = {10.1051/0004-6361/201013990},
       adsurl = {https://ui.adsabs.harvard.edu/abs/2010A&A...513L...7S}
}

@ARTICLE{2024PASP..136l4201B,
       author = {{Blomberg}, Lisa and {El-Badry}, Kareem and {Breivik}, Katelyn and {Caiazzo}, Ilaria and {Nagarajan}, Pranav and {Rodriguez}, Antonio and {van Roestel}, Jan and {Vanderbosch}, Zachary P. and {Yamaguchi}, Natsuko},
        title = "{The Companion Mass Distribution of Post Common Envelope Hot Subdwarf Binaries: Evidence for Boosted and Disrupted Magnetic Braking?}",
      journal = {\pasp},
         year = 2024,
        month = dec,
       volume = {136},
       number = {12},
          eid = {124201},
        pages = {124201},
          doi = {10.1088/1538-3873/ad94a2},
archivePrefix = {arXiv},
       eprint = {2408.15334},
 primaryClass = {astro-ph.SR},
       adsurl = {https://ui.adsabs.harvard.edu/abs/2024PASP..136l4201B}
}

@ARTICLE{2024A&A...682A..33B,
       author = {{Belloni}, Diogo and {Schreiber}, Matthias R. and {Moe}, Maxwell and {El-Badry}, Kareem and {Shen}, Ken J.},
        title = "{Evidence for saturated and disrupted magnetic braking from samples of detached close binaries with M and K dwarfs}",
      journal = {\aap},
         year = 2024,
        month = feb,
       volume = {682},
          eid = {A33},
        pages = {A33},
          doi = {10.1051/0004-6361/202347931},
archivePrefix = {arXiv},
       eprint = {2311.04309},
 primaryClass = {astro-ph.SR},
       adsurl = {https://ui.adsabs.harvard.edu/abs/2024A&A...682A..33B}
}

@ARTICLE{1987ApJ...318..794H,
       author = {{Hjellming}, Michael S. and {Webbink}, Ronald F.},
        title = "{Thresholds for Rapid Mass Transfer in Binary System. I. Polytropic Models}",
      journal = {\apj},
         year = 1987,
        month = jul,
       volume = {318},
        pages = {794},
          doi = {10.1086/165412},
       adsurl = {https://ui.adsabs.harvard.edu/abs/1987ApJ...318..794H}
}

@ARTICLE{2002MNRAS.329..897H,
       author = {{Hurley}, Jarrod R. and {Tout}, Christopher A. and {Pols}, Onno R.},
        title = "{Evolution of binary stars and the effect of tides on binary populations}",
      journal = {\mnras},
         year = 2002,
        month = feb,
       volume = {329},
       number = {4},
        pages = {897-928},
          doi = {10.1046/j.1365-8711.2002.05038.x},
archivePrefix = {arXiv},
       eprint = {astro-ph/0201220},
 primaryClass = {astro-ph},
       adsurl = {https://ui.adsabs.harvard.edu/abs/2002MNRAS.329..897H}
}

@ARTICLE{2010MNRAS.403..179D,
       author = {{Davis}, P.~J. and {Kolb}, U. and {Willems}, B.},
        title = "{A comprehensive population synthesis study of post-common envelope binaries}",
      journal = {\mnras},
         year = 2010,
        month = mar,
       volume = {403},
       number = {1},
        pages = {179-195},
          doi = {10.1111/j.1365-2966.2009.16138.x},
archivePrefix = {arXiv},
       eprint = {0903.4152},
 primaryClass = {astro-ph.SR},
       adsurl = {https://ui.adsabs.harvard.edu/abs/2010MNRAS.403..179D}
}

@ARTICLE{2013A&A...557A..87T,
       author = {{Toonen}, S. and {Nelemans}, G.},
        title = "{The effect of common-envelope evolution on the visible population of post-common-envelope binaries}",
      journal = {\aap},
         year = 2013,
        month = sep,
       volume = {557},
          eid = {A87},
        pages = {A87},
          doi = {10.1051/0004-6361/201321753},
archivePrefix = {arXiv},
       eprint = {1309.0327},
 primaryClass = {astro-ph.SR},
       adsurl = {https://ui.adsabs.harvard.edu/abs/2013A&A...557A..87T}
}

@ARTICLE{2014A&A...568A..68Z,
       author = {{Zorotovic}, M. and {Schreiber}, M.~R. and {Garc{\'\i}a-Berro}, E. and {Camacho}, J. and {Torres}, S. and {Rebassa-Mansergas}, A. and {G{\"a}nsicke}, B.~T.},
        title = "{Monte Carlo simulations of post-common-envelope white dwarf + main sequence binaries: The effects of including recombination energy}",
      journal = {\aap},
         year = 2014,
        month = aug,
       volume = {568},
          eid = {A68},
        pages = {A68},
          doi = {10.1051/0004-6361/201323039},
archivePrefix = {arXiv},
       eprint = {1407.3301},
 primaryClass = {astro-ph.SR},
       adsurl = {https://ui.adsabs.harvard.edu/abs/2014A&A...568A..68Z}
}

@ARTICLE{2016ApJ...826...53A,
       author = {{Ablimit}, Iminhaji and {Maeda}, Keiichi and {Li}, Xiang-Dong},
        title = "{Monte Carlo Population Synthesis of Post-common-envelope White Dwarf Binaries and Type Ia supernova Rate}",
      journal = {\apj},
         year = 2016,
        month = jul,
       volume = {826},
       number = {1},
          eid = {53},
        pages = {53},
          doi = {10.3847/0004-637X/826/1/53},
archivePrefix = {arXiv},
       eprint = {1605.03646},
 primaryClass = {astro-ph.HE},
       adsurl = {https://ui.adsabs.harvard.edu/abs/2016ApJ...826...53A}
}

@ARTICLE{2021MNRAS.501.1677H,
       author = {{Hernandez}, M.~S. and {Schreiber}, M.~R. and {Parsons}, S.~G. and {G{\"a}nsicke}, B.~T. and {Lagos}, F. and {Raddi}, R. and {Toloza}, O. and {Tovmassian}, G. and {Zorotovic}, M. and {Irawati}, P. and {Past{\'e}n}, E. and {Rebassa-Mansergas}, A. and {Ren}, J.~J. and {Rittipruk}, P. and {Tappert}, C.},
        title = "{The White Dwarf Binary Pathways Survey - IV. Three close white dwarf binaries with G-type secondary stars}",
      journal = {\mnras},
         year = 2021,
        month = feb,
       volume = {501},
       number = {2},
        pages = {1677-1689},
          doi = {10.1093/mnras/staa3815},
archivePrefix = {arXiv},
       eprint = {2012.04683},
 primaryClass = {astro-ph.SR},
       adsurl = {https://ui.adsabs.harvard.edu/abs/2021MNRAS.501.1677H}
}

@ARTICLE{2022MNRAS.512.1843H,
       author = {{Hernandez}, M.~S. and {Schreiber}, M.~R. and {Parsons}, S.~G. and {G{\"a}nsicke}, B.~T. and {Toloza}, O. and {Tovmassian}, G. and {Zorotovic}, M. and {Lagos}, F. and {Raddi}, R. and {Rebassa-Mansergas}, A. and {Ren}, J.~J. and {Tappert}, C.},
        title = "{The white dwarf binary pathways survey - VI. Two close post-common envelope binaries with TESS light curves}",
      journal = {\mnras},
         year = 2022,
        month = may,
       volume = {512},
       number = {2},
        pages = {1843-1856},
          doi = {10.1093/mnras/stac604},
archivePrefix = {arXiv},
       eprint = {2203.01745},
 primaryClass = {astro-ph.SR},
       adsurl = {https://ui.adsabs.harvard.edu/abs/2022MNRAS.512.1843H}
}

@ARTICLE{2022MNRAS.517.2867H,
       author = {{Hernandez}, M.~S. and {Schreiber}, M.~R. and {Parsons}, S.~G. and {G{\"a}nsicke}, B.~T. and {Toloza}, O. and {Zorotovic}, M. and {Raddi}, R. and {Rebassa-Mansergas}, A. and {Ren}, J.~J.},
        title = "{The white dwarf binary pathways survey - VIII. A post-common envelope binary with a massive white dwarf and an active G-type secondary star}",
      journal = {\mnras},
         year = 2022,
        month = dec,
       volume = {517},
       number = {2},
        pages = {2867-2875},
          doi = {10.1093/mnras/stac2837},
archivePrefix = {arXiv},
       eprint = {2209.15591},
 primaryClass = {astro-ph.SR},
       adsurl = {https://ui.adsabs.harvard.edu/abs/2022MNRAS.517.2867H}
}

@ARTICLE{2024MNRAS.52711719Y,
       author = {{Yamaguchi}, Natsuko and {El-Badry}, Kareem and {Fuller}, Jim and {Latham}, David W. and {Cargile}, Phillip A. and {Mazeh}, Tsevi and {Shahaf}, Sahar and {Bieryla}, Allyson and {Buchhave}, Lars A. and {Hobson}, Melissa},
        title = "{Wide post-common envelope binaries containing ultramassive white dwarfs: evidence for efficient envelope ejection in massive asymptotic giant branch stars}",
      journal = {\mnras},
         year = 2024,
        month = feb,
       volume = {527},
       number = {4},
        pages = {11719-11739},
          doi = {10.1093/mnras/stad4005},
archivePrefix = {arXiv},
       eprint = {2309.15905},
 primaryClass = {astro-ph.SR},
       adsurl = {https://ui.adsabs.harvard.edu/abs/2024MNRAS.52711719Y}
}

@ARTICLE{2024PASP..136h4202Y,
       author = {{Yamaguchi}, Natsuko and {El-Badry}, Kareem and {Rees}, Natalie R. and {Shahaf}, Sahar and {Mazeh}, Tsevi and {Andrae}, Ren{\'r}},
        title = "{Wide Post-common Envelope Binaries from Gaia: Orbit Validation and Formation Models}",
      journal = {\pasp},
         year = 2024,
        month = aug,
       volume = {136},
       number = {8},
          eid = {084202},
        pages = {084202},
          doi = {10.1088/1538-3873/ad6809},
archivePrefix = {arXiv},
       eprint = {2405.06020},
 primaryClass = {astro-ph.SR},
       adsurl = {https://ui.adsabs.harvard.edu/abs/2024PASP..136h4202Y}
}

@ARTICLE{2026AJ....171..159M,
       author = {{Motherway}, Erin and {Linck}, Evan and {Mathieu}, Robert D. and {Dixon}, Don and {Stassun}, Keivan G. and {Breivik}, Katelyn and {Majewski}, Steve and {Pols}, Onno R.},
        title = "{A Not-so-compact Companion: Massive, Oversized White Dwarf in a Post-common-envelope Eclipsing Binary}",
      journal = {\aj},
         year = 2026,
        month = mar,
       volume = {171},
       number = {3},
          eid = {159},
        pages = {159},
          doi = {10.3847/1538-3881/ae3b42},
archivePrefix = {arXiv},
       eprint = {2601.14378},
 primaryClass = {astro-ph.SR},
       adsurl = {https://ui.adsabs.harvard.edu/abs/2026AJ....171..159M}
}

@ARTICLE{2024A&A...686A..61B,
       author = {{Belloni}, Diogo and {Zorotovic}, Monica and {Schreiber}, Matthias R. and {Parsons}, Steven G. and {Moe}, Maxwell and {Garbutt}, James A.},
        title = "{Formation of long-period post-common envelope binaries. I. No extra energy is needed to explain oxygen-neon white dwarfs paired with AFGK-type main-sequence stars}",
      journal = {\aap},
         year = 2024,
        month = jun,
       volume = {686},
          eid = {A61},
        pages = {A61},
          doi = {10.1051/0004-6361/202449235},
archivePrefix = {arXiv},
       eprint = {2401.07692},
 primaryClass = {astro-ph.SR},
       adsurl = {https://ui.adsabs.harvard.edu/abs/2024A&A...686A..61B}
}

@ARTICLE{2024A&A...687A..12B,
       author = {{Belloni}, Diogo and {Schreiber}, Matthias R. and {Zorotovic}, Monica},
        title = "{Formation of long-period post-common-envelope binaries. II. Explaining the self-lensing binary KOI 3278}",
      journal = {\aap},
         year = 2024,
        month = jul,
       volume = {687},
          eid = {A12},
        pages = {A12},
          doi = {10.1051/0004-6361/202449320},
archivePrefix = {arXiv},
       eprint = {2401.17510},
 primaryClass = {astro-ph.SR},
       adsurl = {https://ui.adsabs.harvard.edu/abs/2024A&A...687A..12B}
}

@ARTICLE{2010ApJ...717..724G,
       author = {{Ge}, Hongwei and {Hjellming}, Michael S. and {Webbink}, Ronald F. and {Chen}, Xuefei and {Han}, Zhanwen},
        title = "{Adiabatic Mass Loss in Binary Stars. I. Computational Method}",
      journal = {\apj},
         year = 2010,
        month = jul,
       volume = {717},
       number = {2},
        pages = {724-738},
          doi = {10.1088/0004-637X/717/2/724},
archivePrefix = {arXiv},
       eprint = {1005.3099},
 primaryClass = {astro-ph.SR},
       adsurl = {https://ui.adsabs.harvard.edu/abs/2010ApJ...717..724G}
}

@ARTICLE{2015ApJ...812...40G,
       author = {{Ge}, Hongwei and {Webbink}, Ronald F. and {Chen}, Xuefei and {Han}, Zhanwen},
        title = "{Adiabatic Mass Loss in Binary Stars. II. From Zero-age Main Sequence to the Base of the Giant Branch}",
      journal = {\apj},
         year = 2015,
        month = oct,
       volume = {812},
       number = {1},
          eid = {40},
        pages = {40},
          doi = {10.1088/0004-637X/812/1/40},
archivePrefix = {arXiv},
       eprint = {1507.04843},
 primaryClass = {astro-ph.SR},
       adsurl = {https://ui.adsabs.harvard.edu/abs/2015ApJ...812...40G}
}

@ARTICLE{2020ApJ...899..132G,
       author = {{Ge}, Hongwei and {Webbink}, Ronald F. and {Chen}, Xuefei and {Han}, Zhanwen},
        title = "{Adiabatic Mass Loss in Binary Stars. III. From the Base of the Red Giant Branch to the Tip of the Asymptotic Giant Branch}",
      journal = {\apj},
         year = 2020,
        month = aug,
       volume = {899},
       number = {2},
          eid = {132},
        pages = {132},
          doi = {10.3847/1538-4357/aba7b7},
archivePrefix = {arXiv},
       eprint = {2007.09848},
 primaryClass = {astro-ph.SR},
       adsurl = {https://ui.adsabs.harvard.edu/abs/2020ApJ...899..132G}
}

@ARTICLE{2023ApJ...945....7G,
       author = {{Ge}, Hongwei and {Tout}, Christopher A. and {Chen}, Xuefei and {Sarkar}, Arnab and {Walton}, Dominic J. and {Han}, Zhanwen},
        title = "{Criteria for Dynamical Timescale Mass Transfer of Metal-poor Intermediate-mass Stars}",
      journal = {\apj},
         year = 2023,
        month = mar,
       volume = {945},
       number = {1},
          eid = {7},
        pages = {7},
          doi = {10.3847/1538-4357/acb7e9},
archivePrefix = {arXiv},
       eprint = {2302.00183},
 primaryClass = {astro-ph.SR},
       adsurl = {https://ui.adsabs.harvard.edu/abs/2023ApJ...945....7G}
}

@ARTICLE{2010ApJ...716..114X,
       author = {{Xu}, Xiao-Jie and {Li}, Xiang-Dong},
        title = "{On the Binding Energy Parameter {\ensuremath{\lambda}} of Common Envelope Evolution}",
      journal = {\apj},
         year = 2010,
        month = jun,
       volume = {716},
       number = {1},
        pages = {114-121},
          doi = {10.1088/0004-637X/716/1/114},
archivePrefix = {arXiv},
       eprint = {1004.4957},
 primaryClass = {astro-ph.SR},
       adsurl = {https://ui.adsabs.harvard.edu/abs/2010ApJ...716..114X}
}

@ARTICLE{2010ApJ...722.1985X,
       author = {{Xu}, Xiao-Jie and {Li}, Xiang-Dong},
        title = "{ERRATUM: ``On the Binding Energy Parameter {\ensuremath{\lambda}} of Common Envelope Evolution'' <A href=``/abs/2010ApJ...716..114X''>(2010, ApJ, 716, 114)</A>}",
      journal = {\apj},
         year = 2010,
        month = oct,
       volume = {722},
       number = {2},
        pages = {1985-1988},
          doi = {10.1088/0004-637X/722/2/1985},
       adsurl = {https://ui.adsabs.harvard.edu/abs/2010ApJ...722.1985X}
}

@ARTICLE{2000MNRAS.315..543H,
       author = {{Hurley}, Jarrod R. and {Pols}, Onno R. and {Tout}, Christopher A.},
        title = "{Comprehensive analytic formulae for stellar evolution as a function of mass and metallicity}",
      journal = {\mnras},
         year = 2000,
        month = jul,
       volume = {315},
       number = {3},
        pages = {543-569},
          doi = {10.1046/j.1365-8711.2000.03426.x},
archivePrefix = {arXiv},
       eprint = {astro-ph/0001295},
 primaryClass = {astro-ph},
       adsurl = {https://ui.adsabs.harvard.edu/abs/2000MNRAS.315..543H}
}

@ARTICLE{1979ApJS...41..513M,
       author = {{Miller}, G.~E. and {Scalo}, J.~M.},
        title = "{The Initial Mass Function and Stellar Birthrate in the Solar Neighborhood}",
      journal = {\apjs},
         year = 1979,
        month = nov,
       volume = {41},
        pages = {513},
          doi = {10.1086/190629},
       adsurl = {https://ui.adsabs.harvard.edu/abs/1979ApJS...41..513M}
}

@ARTICLE{1989ApJ...347..998E,
       author = {{Eggleton}, Peter P. and {Fitchett}, Michael J. and {Tout}, Christopher A.},
        title = "{The Distribution of Visual Binaries with Two Bright Components}",
      journal = {\apj},
         year = 1989,
        month = dec,
       volume = {347},
        pages = {998},
          doi = {10.1086/168190},
       adsurl = {https://ui.adsabs.harvard.edu/abs/1989ApJ...347..998E}
}

@ARTICLE{2020RAA....20..161H,
       author = {{Han}, Zhan-Wen and {Ge}, Hong-Wei and {Chen}, Xue-Fei and {Chen}, Hai-Liang},
        title = "{Binary Population Synthesis}",
      journal = {Research in Astronomy and Astrophysics},
         year = 2020,
        month = oct,
       volume = {20},
       number = {10},
          eid = {161},
        pages = {161},
          doi = {10.1088/1674-4527/20/10/161},
archivePrefix = {arXiv},
       eprint = {2009.08611},
 primaryClass = {astro-ph.SR},
       adsurl = {https://ui.adsabs.harvard.edu/abs/2020RAA....20..161H}
}

@ARTICLE{2013A&A...549A..95Z,
       author = {{Zorotovic}, M. and {Schreiber}, M.~R.},
        title = "{Origin of apparent period variations in eclipsing post-common-envelope binaries}",
      journal = {\aap},
         year = 2013,
        month = jan,
       volume = {549},
          eid = {A95},
        pages = {A95},
          doi = {10.1051/0004-6361/201220321},
archivePrefix = {arXiv},
       eprint = {1211.5356},
 primaryClass = {astro-ph.SR},
       adsurl = {https://ui.adsabs.harvard.edu/abs/2013A&A...549A..95Z}
}

@ARTICLE{2008MNRAS.389.1563D,
       author = {{Davis}, P.~J. and {Kolb}, U. and {Willems}, B. and {G{\"a}nsicke}, B.~T.},
        title = "{How many cataclysmic variables are crossing the period gap? A test for the disruption of magnetic braking}",
      journal = {\mnras},
         year = 2008,
        month = oct,
       volume = {389},
       number = {4},
        pages = {1563-1576},
          doi = {10.1111/j.1365-2966.2008.13675.x},
archivePrefix = {arXiv},
       eprint = {0805.4700},
 primaryClass = {astro-ph},
       adsurl = {https://ui.adsabs.harvard.edu/abs/2008MNRAS.389.1563D}
}

@ARTICLE{2020ApJS..249....9G,
       author = {{Ge}, Hongwei and {Webbink}, Ronald F. and {Han}, Zhanwen},
        title = "{The Thermal Equilibrium Mass-loss Model and Its Applications in Binary Evolution}",
      journal = {\apjs},
         year = 2020,
        month = jul,
       volume = {249},
       number = {1},
          eid = {9},
        pages = {9},
          doi = {10.3847/1538-4365/ab98f6},
archivePrefix = {arXiv},
       eprint = {2006.00774},
 primaryClass = {astro-ph.SR},
       adsurl = {https://ui.adsabs.harvard.edu/abs/2020ApJS..249....9G}
}

@ARTICLE{2023A&A...669A..82L,
       author = {{Li}, Zhenwei and {Chen}, Xuefei and {Ge}, Hongwei and {Chen}, Hai-Liang and {Han}, Zhanwen},
        title = "{Influence of a mass transfer stability criterion on double white dwarf populations}",
      journal = {\aap},
         year = 2023,
        month = jan,
       volume = {669},
          eid = {A82},
        pages = {A82},
          doi = {10.1051/0004-6361/202243893},
archivePrefix = {arXiv},
       eprint = {2211.01861},
 primaryClass = {astro-ph.SR},
       adsurl = {https://ui.adsabs.harvard.edu/abs/2023A&A...669A..82L}
}

@ARTICLE{1984ApJ...277..355W,
       author = {{Webbink}, R.~F.},
        title = "{Double white dwarfs as progenitors of R Coronae Borealis stars and type I supernovae.}",
      journal = {\apj},
         year = 1984,
        month = feb,
       volume = {277},
        pages = {355-360},
          doi = {10.1086/161701},
       adsurl = {https://ui.adsabs.harvard.edu/abs/1984ApJ...277..355W}
}

@ARTICLE{1988ApJ...329..764L,
       author = {{Livio}, Mario and {Soker}, Noam},
        title = "{The Common Envelope Phase in the Evolution of Binary Stars}",
      journal = {\apj},
         year = 1988,
        month = jun,
       volume = {329},
        pages = {764},
          doi = {10.1086/166419},
       adsurl = {https://ui.adsabs.harvard.edu/abs/1988ApJ...329..764L}
}

@ARTICLE{2022ApJ...933..137G,
       author = {{Ge}, Hongwei and {Tout}, Christopher A. and {Chen}, Xuefei and {Kruckow}, Matthias U. and {Chen}, Hailiang and {Jiang}, Dengkai and {Li}, Zhenwei and {Liu}, Zhengwei and {Han}, Zhanwen},
        title = "{The Common Envelope Evolution Outcome-A Case Study on Hot Subdwarf B Stars}",
      journal = {\apj},
         year = 2022,
        month = jul,
       volume = {933},
       number = {2},
          eid = {137},
        pages = {137},
          doi = {10.3847/1538-4357/ac75d3},
archivePrefix = {arXiv},
       eprint = {2205.14256},
 primaryClass = {astro-ph.SR},
       adsurl = {https://ui.adsabs.harvard.edu/abs/2022ApJ...933..137G}
}

@ARTICLE{2024ApJ...961..202G,
       author = {{Ge}, Hongwei and {Tout}, Christopher A. and {Webbink}, Ronald F. and {Chen}, Xuefei and {Sarkar}, Arnab and {Li}, Jiao and {Li}, Zhenwei and {Zhang}, Lifu and {Han}, Zhanwen},
        title = "{The Common Envelope Evolution Outcome. II. Short-orbital-period Hot Subdwarf B Binaries Reveal a Clear Picture}",
      journal = {\apj},
         year = 2024,
        month = feb,
       volume = {961},
       number = {2},
          eid = {202},
        pages = {202},
          doi = {10.3847/1538-4357/ad158e},
archivePrefix = {arXiv},
       eprint = {2311.17304},
 primaryClass = {astro-ph.SR},
       adsurl = {https://ui.adsabs.harvard.edu/abs/2024ApJ...961..202G}
}

@ARTICLE{2022ApJS..258...34R,
       author = {{Riley}, Jeff and {Agrawal}, Poojan and {Barrett}, Jim W. and {Boyett}, Kristan N.~K. and {Broekgaarden}, Floor S. and {Chattopadhyay}, Debatri and {Gaebel}, Sebastian M. and {Gittins}, Fabian and {Hirai}, Ryosuke and {Howitt}, George and {Justham}, Stephen and {Khandelwal}, Lokesh and {Kummer}, Floris and {Lau}, Mike Y.~M. and {Mandel}, Ilya and {de Mink}, Selma E. and {Neijssel}, Coenraad and {Riley}, Tim and {van Son}, Lieke and {Stevenson}, Simon and {Vigna-G{\'o}mez}, Alejandro and {Vinciguerra}, Serena and {Wagg}, Tom and {Willcox}, Reinhold and {Team Compas}},
        title = "{Rapid Stellar and Binary Population Synthesis with COMPAS}",
      journal = {\apjs},
         year = 2022,
        month = feb,
       volume = {258},
       number = {2},
          eid = {34},
        pages = {34},
          doi = {10.3847/1538-4365/ac416c},
archivePrefix = {arXiv},
       eprint = {2109.10352},
 primaryClass = {astro-ph.IM},
       adsurl = {https://ui.adsabs.harvard.edu/abs/2022ApJS..258...34R}
}

@ARTICLE{2023MNRAS.518.4579P,
       author = {{Parsons}, S.~G. and {Hernandez}, M.~S. and {Toloza}, O. and {Zorotovic}, M. and {Schreiber}, M.~R. and {G{\"a}nsicke}, B.~T. and {Lagos}, F. and {Raddi}, R. and {Rebassa-Mansergas}, A. and {Ren}, J.~J. and {Koester}, D.},
        title = "{The white dwarf binary pathways survey - IX. Three long period white dwarf plus subgiant binaries}",
      journal = {\mnras},
         year = 2023,
        month = jan,
       volume = {518},
       number = {3},
        pages = {4579-4594},
          doi = {10.1093/mnras/stac3368},
archivePrefix = {arXiv},
       eprint = {2211.08440},
 primaryClass = {astro-ph.SR},
       adsurl = {https://ui.adsabs.harvard.edu/abs/2023MNRAS.518.4579P}
}

@ARTICLE{1995MNRAS.273..731R,
       author = {{Rappaport}, S. and {Podsiadlowski}, Ph. and {Joss}, P.~C. and {Di Stefano}, R. and {Han}, Z.},
        title = "{The relation between white dwarf mass and orbital period in wide binary radio pulsars}",
      journal = {\mnras},
         year = 1995,
        month = apr,
       volume = {273},
       number = {3},
        pages = {731-741},
          doi = {10.1093/mnras/273.3.731},
       adsurl = {https://ui.adsabs.harvard.edu/abs/1995MNRAS.273..731R}
}

@ARTICLE{2023MNRAS.521.1880B,
       author = {{Brown}, Alex J. and {Parsons}, Steven G. and {van Roestel}, Jan and {Rebassa-Mansergas}, Alberto and {Breedt}, Elm{\'e} and {Dhillon}, Vik S. and {Dyer}, Martin J. and {Green}, Matthew J. and {Kerry}, Paul and {Littlefair}, Stuart P. and {Marsh}, Thomas R. and {Munday}, James and {Pelisoli}, Ingrid and {Sahman}, David I. and {Wild}, James F.},
        title = "{Photometric follow-up of 43 new eclipsing white dwarf plus main-sequence binaries from the ZTF survey}",
      journal = {\mnras},
         year = 2023,
        month = may,
       volume = {521},
       number = {2},
        pages = {1880-1896},
          doi = {10.1093/mnras/stad612},
archivePrefix = {arXiv},
       eprint = {2302.11392},
 primaryClass = {astro-ph.SR},
       adsurl = {https://ui.adsabs.harvard.edu/abs/2023MNRAS.521.1880B}
}

@ARTICLE{2012MNRAS.423..320R,
       author = {{Rebassa-Mansergas}, A. and {Zorotovic}, M. and {Schreiber}, M.~R. and {G{\"a}nsicke}, B.~T. and {Southworth}, J. and {Nebot G{\'o}mez-Mor{\'a}n}, A. and {Tappert}, C. and {Koester}, D. and {Pyrzas}, S. and {Papadaki}, C. and {Schmidtobreick}, L. and {Schwope}, A. and {Toloza}, O.},
        title = "{Post-common envelope binaries from SDSS - XVI. Long orbital period systems and the energy budget of common envelope evolution}",
      journal = {\mnras},
         year = 2012,
        month = jun,
       volume = {423},
       number = {1},
        pages = {320-327},
          doi = {10.1111/j.1365-2966.2012.20880.x},
archivePrefix = {arXiv},
       eprint = {1203.1208},
 primaryClass = {astro-ph.SR},
       adsurl = {https://ui.adsabs.harvard.edu/abs/2012MNRAS.423..320R}
}

@ARTICLE{2021ApJ...920...86K,
       author = {{Kruckow}, Matthias U. and {Neunteufel}, Patrick G. and {Di Stefano}, Rosanne and {Gao}, Yan and {Kobayashi}, Chiaki},
        title = "{A Catalog of Potential Post-Common Envelope Binaries}",
      journal = {\apj},
         year = 2021,
        month = oct,
       volume = {920},
       number = {2},
          eid = {86},
        pages = {86},
          doi = {10.3847/1538-4357/ac13ac},
archivePrefix = {arXiv},
       eprint = {2107.05221},
 primaryClass = {astro-ph.SR},
       adsurl = {https://ui.adsabs.harvard.edu/abs/2021ApJ...920...86K}
}

@ARTICLE{2025A&A...698A..81L,
       author = {{Liu}, Qichun and {Wang}, Xiaofeng and {Lin}, Jie and {Wu}, Chengyuan and {Li}, Chunqian and {Filippenko}, Alexei V. and {Brink}, Thomas G. and {Yang}, Yi and {Zheng}, Weikang and {Liu}, Cheng and {Song}, Cuiying and {Kovalev}, Mikhail and {Ge}, Hongwei and {Zhang}, Fenghui and {Zhang}, Xiaobin and {Xia}, Qiqi and {Peng}, Haowei and {Xi}, Gaobo and {Mo}, Jun and {Yan}, Shengyu and {Shi}, Jianrong and {Li}, Jiangdan and {Yi}, Tuan},
        title = "{A post-common-envelope binary with double-peaked Balmer emission lines from TMTS}",
      journal = {\aap},
         year = 2025,
        month = jun,
       volume = {698},
          eid = {A81},
        pages = {A81},
          doi = {10.1051/0004-6361/202553732},
archivePrefix = {arXiv},
       eprint = {2504.18202},
 primaryClass = {astro-ph.SR},
       adsurl = {https://ui.adsabs.harvard.edu/abs/2025A&A...698A..81L}
}

@ARTICLE{2009MNRAS.394..978P,
       author = {{Pyrzas}, S. and {G{\"a}nsicke}, B.~T. and {Marsh}, T.~R. and {Aungwerojwit}, A. and {Rebassa-Mansergas}, A. and {Rodr{\'\i}guez-Gil}, P. and {Southworth}, J. and {Schreiber}, M.~R. and {Nebot Gomez-Moran}, A. and {Koester}, D.},
        title = "{Post-common-envelope binaries from SDSS - V. Four eclipsing white dwarf main-sequence binaries}",
      journal = {\mnras},
         year = 2009,
        month = apr,
       volume = {394},
       number = {2},
        pages = {978-994},
          doi = {10.1111/j.1365-2966.2008.14378.x},
archivePrefix = {arXiv},
       eprint = {0812.2510},
 primaryClass = {astro-ph},
       adsurl = {https://ui.adsabs.harvard.edu/abs/2009MNRAS.394..978P}
}

@ARTICLE{2024MNRAS.532.1718Q,
       author = {{Qi}, Senyu and {Gu}, Wei-Min and {Zhang}, Zhi-Xiang and {Yi}, Tuan and {Liu}, Jin-Zhong and {Zheng}, Ling-Lin},
        title = "{Two dynamically discovered compact object candidate binary systems from LAMOST low-resolution survey}",
      journal = {\mnras},
         year = 2024,
        month = aug,
       volume = {532},
       number = {2},
        pages = {1718-1728},
          doi = {10.1093/mnras/stae1590},
archivePrefix = {arXiv},
       eprint = {2406.17340},
 primaryClass = {astro-ph.SR},
       adsurl = {https://ui.adsabs.harvard.edu/abs/2024MNRAS.532.1718Q}
}

@ARTICLE{2020MNRAS.493.5208K,
       author = {{Krushinsky}, Vadim and {Benni}, Paul and {Burdanov}, Artem and {Antokhin}, Igor and {Antokhina}, Eleonora and {Jehin}, Emmanu{\"e}l and {Barkaoui}, Khalid and {Fitzsimmons}, Alan and {Gibson}, Christopher and {Gillon}, Micha{\"e}l and {Popov}, Alexander and {Ba{\textcommabelow s}t{\"u}rk}, {\"O}zg{\"u}r and {Benkhaldoun}, Zouhair and {Marchini}, Alessandro and {Papini}, Riccardo and {Salvaggio}, Fabio and {Brazhko}, Varvara},
        title = "{Discovery of a pre-cataclysmic binary with unusual chromaticity of the eclipsed white dwarf by the GPX survey}",
      journal = {\mnras},
         year = 2020,
        month = apr,
       volume = {493},
       number = {4},
        pages = {5208-5217},
          doi = {10.1093/mnras/staa547},
archivePrefix = {arXiv},
       eprint = {2002.08031},
 primaryClass = {astro-ph.SR},
       adsurl = {https://ui.adsabs.harvard.edu/abs/2020MNRAS.493.5208K}
}

@ARTICLE{2025ApJ...984...42H,
       author = {{He}, Yuji and {Yuan}, Hailong and {Bai}, Zhongrui and {Yang}, Mingkuan and {Wang}, Mengxin and {Dong}, Yiqiao and {Huang}, Xin and {Zhou}, Ming and {Liu}, Qian and {Yang}, Xiaozhen and {Li}, Ganyu and {Jiang}, Ziyue and {Zhang}, Haotong},
        title = "{LAMOST J101356.33+272410.7: A Detached White Dwarf─Main-sequence Binary with a Massive White Dwarf within the Period Gap}",
      journal = {\apj},
         year = 2025,
        month = may,
       volume = {984},
       number = {1},
          eid = {42},
        pages = {42},
          doi = {10.3847/1538-4357/adc0fa},
archivePrefix = {arXiv},
       eprint = {2501.01171},
 primaryClass = {astro-ph.SR},
       adsurl = {https://ui.adsabs.harvard.edu/abs/2025ApJ...984...42H}
}

@ARTICLE{2026PASJ...78..382S,
       author = {{Shiraishi}, Yuta and {Hotokezaka}, Kenta and {Masuda}, Kento and {Honda}, Satoshi and {Tanikawa}, Ataru and {Janssens}, Soetkin and {Tokuno}, Takato and {Shimasue}, Takumi and {Honjo}, Ryoga and {Sato}, Bun'ei and {Omiya}, Masashi and {Tajitsu}, Akito and {Izumiura}, Hideyuki},
        title = "{Two unseen massive white dwarf candidates in close binaries}",
      journal = {\pasj},
         year = 2026,
        month = apr,
       volume = {78},
       number = {2},
        pages = {382-404},
          doi = {10.1093/pasj/psaf148},
archivePrefix = {arXiv},
       eprint = {2509.12808},
 primaryClass = {astro-ph.SR},
       adsurl = {https://ui.adsabs.harvard.edu/abs/2026PASJ...78..382S}
}

@ARTICLE{2018AJ....155..144K,
       author = {{Kawahara}, Hajime and {Masuda}, Kento and {MacLeod}, Morgan and {Latham}, David W. and {Bieryla}, Allyson and {Benomar}, Othman},
        title = "{Discovery of Three Self-lensing Binaries from Kepler}",
      journal = {\aj},
         year = 2018,
        month = mar,
       volume = {155},
       number = {3},
          eid = {144},
        pages = {144},
          doi = {10.3847/1538-3881/aaaaaf},
archivePrefix = {arXiv},
       eprint = {1801.07874},
 primaryClass = {astro-ph.SR},
       adsurl = {https://ui.adsabs.harvard.edu/abs/2018AJ....155..144K}
}

@ARTICLE{2023MNRAS.520.3187S,
       author = {{Sarkar}, Arnab and {Ge}, Hongwei and {Tout}, Christopher A.},
        title = "{Evolved cataclysmic variables as progenitors of AM CVn stars}",
      journal = {\mnras},
         year = 2023,
        month = apr,
       volume = {520},
       number = {2},
        pages = {3187-3200},
          doi = {10.1093/mnras/stad354},
archivePrefix = {arXiv},
       eprint = {2301.12992},
 primaryClass = {astro-ph.SR},
       adsurl = {https://ui.adsabs.harvard.edu/abs/2023MNRAS.520.3187S}
}

@ARTICLE{2023A&A...678A..34B,
       author = {{Belloni}, Diogo and {Schreiber}, Matthias R.},
        title = "{Reversing the verdict: Cataclysmic variables could be the dominant progenitors of AM CVn binaries after all}",
      journal = {\aap},
         year = 2023,
        month = oct,
       volume = {678},
          eid = {A34},
        pages = {A34},
          doi = {10.1051/0004-6361/202347047},
archivePrefix = {arXiv},
       eprint = {2305.19312},
 primaryClass = {astro-ph.SR},
       adsurl = {https://ui.adsabs.harvard.edu/abs/2023A&A...678A..34B}
}

@ARTICLE{2025A&A...703A.119T,
       author = {{Tovmassian}, Gagik and {Belloni}, Diogo and {Pala}, Anna F. and {Kupfer}, Thomas and {Yu}, Weitian and {G{\"a}nsicke}, Boris T. and {Waagen}, Elizabeth O. and {Gonz{\'a}lez-Carballo}, Juan-Luis and {Szkody}, Paula and {de Martino}, Domitilla and {Schreiber}, Matthias R. and {Long}, Knox S. and {Bedard}, Alan and {Bednarz}, Slawomir and {Berenguer}, Jordi and {Bernacki}, Krzysztof and {Bolzoni}, Simone and {Botana-Alb{\'a}}, Carlos and {Cantrell}, Christopher and {Cooney}, Walt and {Cynamon}, Charles and {De la Fuente Fern{\'a}ndez}, Pablo and {Dufoer}, Sjoerd and {Ma{\~n}anes}, Esteban Fern{\'a}ndez and {Garc{\'\i}a-Cuesta}, Faustino and {Farf{\'a}n}, Rafael Gonzalez and {Fleurant}, Pierre A. and {G{\'o}mez}, Enrique A. and {Green}, Matthew J. and {Hambsch}, Franz-Josef and {Jordanov}, Penko and {Kardasis}, Emmanuel and {Lane}, David and {Lee}, Darrell and {Lima}, Isabel J. and {Mart{\'\i}nez}, Fernando Lim{\'o}n and {Locatelli}, Gianpiero and {Martin-Velasco}, Jose-Luis and {Mendicini}, Daniel J. and {Michaud}, Michel and {Ort{\'\i}z}, Mois{\'e}s Montero Reyes and {Aimar}, Mario Morales and {Myers}, Gordon and {Nogues}, Ramon Naves and {Pappa}, Giuseppe and {Pearce}, Andrew and {Pierce}, James and {Popowicz}, Adam and {Rodrigues}, Claudia V. and {Rodr{\'\i}guez}, Nieves C. and {Amat}, David Quiles and {Reina-Lorenz}, Esteban and {Salto-Gonz{\'a}lez}, Jos{\'e}-Luis and {Shears}, Jeremy and {Sikora}, John and {Steenkamp}, Andr{\'e} and {Stubbings}, Rod and {Young}, Brad and {Walton}, Ivan L.},
        title = "{Revisiting the extremely long-period cataclysmic variables V479 Andromedae and V1082 Sagittarii}",
      journal = {\aap},
         year = 2025,
        month = nov,
       volume = {703},
          eid = {A119},
        pages = {A119},
          doi = {10.1051/0004-6361/202556385},
archivePrefix = {arXiv},
       eprint = {2508.21358},
 primaryClass = {astro-ph.SR},
       adsurl = {https://ui.adsabs.harvard.edu/abs/2025A&A...703A.119T}
}

@ARTICLE{2019ApJ...886L..31V,
       author = {{Van}, Kenny X. and {Ivanova}, Natalia},
        title = "{Evolving LMXBs: CARB Magnetic Braking}",
      journal = {\apjl},
         year = 2019,
        month = dec,
       volume = {886},
       number = {2},
          eid = {L31},
        pages = {L31},
          doi = {10.3847/2041-8213/ab571c},
archivePrefix = {arXiv},
       eprint = {1911.05790},
 primaryClass = {astro-ph.SR},
       adsurl = {https://ui.adsabs.harvard.edu/abs/2019ApJ...886L..31V}
}

@ARTICLE{2025ApJ...988..102G,
       author = {{Gossage}, Seth and {Kiman}, Rocio and {Monsch}, Kristina and {Medina}, Amber A. and {Drake}, Jeremy J. and {Garraffo}, Cecilia and {Lu}, Yuxi(Lucy) and {Wing}, Joshua D. and {Wright}, Nicholas J.},
        title = "{On Convective Turnover Times and Dynamos in Low-mass Stars}",
      journal = {\apj},
         year = 2025,
        month = jul,
       volume = {988},
       number = {1},
          eid = {102},
        pages = {102},
          doi = {10.3847/1538-4357/adde4d},
archivePrefix = {arXiv},
       eprint = {2410.20000},
 primaryClass = {astro-ph.SR},
       adsurl = {https://ui.adsabs.harvard.edu/abs/2025ApJ...988..102G}
}

@ARTICLE{2019ApJ...881...88F,
       author = {{Fleming}, David P. and {Barnes}, Rory and {Davenport}, James R.~A. and {Luger}, Rodrigo},
        title = "{Rotation Period Evolution in Low-mass Binary Stars: The Impact of Tidal Torques and Magnetic Braking}",
      journal = {\apj},
         year = 2019,
        month = aug,
       volume = {881},
       number = {2},
          eid = {88},
        pages = {88},
          doi = {10.3847/1538-4357/ab2ed2},
archivePrefix = {arXiv},
       eprint = {1903.05686},
 primaryClass = {astro-ph.SR},
       adsurl = {https://ui.adsabs.harvard.edu/abs/2019ApJ...881...88F}
}

@ARTICLE{2024ApJ...971...80S,
       author = {{Sun}, Meng and {Gossage}, Seth and {Leiner}, Emily M. and {Geller}, Aaron M.},
        title = "{Stellar Spin-down in Post-mass-transfer Binary Systems}",
      journal = {\apj},
         year = 2024,
        month = aug,
       volume = {971},
       number = {1},
          eid = {80},
        pages = {80},
          doi = {10.3847/1538-4357/ad54be},
archivePrefix = {arXiv},
       eprint = {2403.17279},
 primaryClass = {astro-ph.SR},
       adsurl = {https://ui.adsabs.harvard.edu/abs/2024ApJ...971...80S}
}

@ARTICLE{2025A&A...695A.161S,
       author = {{Santos-Garc{\'\i}a}, A. and {Torres}, S. and {Rebassa-Mansergas}, A. and {Brown}, A.~J.},
        title = "{A population synthesis study of the Gaia 100 pc unresolved white dwarf-main-sequence binary population}",
      journal = {\aap},
         year = 2025,
        month = mar,
       volume = {695},
          eid = {A161},
        pages = {A161},
          doi = {10.1051/0004-6361/202452989},
archivePrefix = {arXiv},
       eprint = {2502.11593},
 primaryClass = {astro-ph.SR},
       adsurl = {https://ui.adsabs.harvard.edu/abs/2025A&A...695A.161S}
}

@ARTICLE{2012MNRAS.419..806R,
       author = {{Rebassa-Mansergas}, A. and {Nebot G{\'o}mez-Mor{\'a}n}, A. and {Schreiber}, M.~R. and {G{\"a}nsicke}, B.~T. and {Schwope}, A. and {Gallardo}, J. and {Koester}, D.},
        title = "{Post-common envelope binaries from SDSS - XIV. The DR7 white dwarf-main-sequence binary catalogue}",
      journal = {\mnras},
         year = 2012,
        month = jan,
       volume = {419},
       number = {1},
        pages = {806-816},
          doi = {10.1111/j.1365-2966.2011.19923.x},
archivePrefix = {arXiv},
       eprint = {1110.1000},
 primaryClass = {astro-ph.SR},
       adsurl = {https://ui.adsabs.harvard.edu/abs/2012MNRAS.419..806R}
}

@ARTICLE{2013ARA&A..51..269D,
       author = {{Duch{\^e}ne}, Gaspard and {Kraus}, Adam},
        title = "{Stellar Multiplicity}",
      journal = {\araa},
         year = 2013,
        month = aug,
       volume = {51},
       number = {1},
        pages = {269-310},
          doi = {10.1146/annurev-astro-081710-102602},
archivePrefix = {arXiv},
       eprint = {1303.3028},
 primaryClass = {astro-ph.SR},
       adsurl = {https://ui.adsabs.harvard.edu/abs/2013ARA&A..51..269D}
}

@ARTICLE{2017ApJS..230...15M,
       author = {{Moe}, Maxwell and {Di Stefano}, Rosanne},
        title = "{Mind Your Ps and Qs: The Interrelation between Period (P) and Mass-ratio (Q) Distributions of Binary Stars}",
      journal = {\apjs},
         year = 2017,
        month = jun,
       volume = {230},
       number = {2},
          eid = {15},
        pages = {15},
          doi = {10.3847/1538-4365/aa6fb6},
archivePrefix = {arXiv},
       eprint = {1606.05347},
 primaryClass = {astro-ph.SR},
       adsurl = {https://ui.adsabs.harvard.edu/abs/2017ApJS..230...15M}
}

@INPROCEEDINGS{1976IAUS...73...75P,
       author = {{Paczynski}, B.},
        title = "{Common Envelope Binaries}",
    booktitle = {Structure and Evolution of Close Binary Systems},
         year = 1976,
       editor = {{Eggleton}, Peter and {Mitton}, Simon and {Whelan}, John},
       series = {IAU Symposium},
       volume = {73},
        month = jan,
        pages = {75},
       adsurl = {https://ui.adsabs.harvard.edu/abs/1976IAUS...73...75P}
}

@ARTICLE{2013A&ARv..21...59I,
       author = {{Ivanova}, N. and {Justham}, S. and {Chen}, X. and {De Marco}, O. and {Fryer}, C.~L. and {Gaburov}, E. and {Ge}, H. and {Glebbeek}, E. and {Han}, Z. and {Li}, X.-D. and {Lu}, G. and {Marsh}, T. and {Podsiadlowski}, P. and {Potter}, A. and {Soker}, N. and {Taam}, R. and {Tauris}, T.~M. and {van den Heuvel}, E.~P.~J. and {Webbink}, R.~F.},
        title = "{Common envelope evolution: where we stand and how we can move forward}",
      journal = {\aapr},
         year = 2013,
        month = feb,
       volume = {21},
          eid = {59},
        pages = {59},
          doi = {10.1007/s00159-013-0059-2},
archivePrefix = {arXiv},
       eprint = {1209.4302},
 primaryClass = {astro-ph.HE},
       adsurl = {https://ui.adsabs.harvard.edu/abs/2013A&ARv..21...59I}
}

@ARTICLE{2026PASP..138c4202S,
       author = {{Shariat}, Cheyanne and {El-Badry}, Kareem},
        title = "{A Global View of Post-interaction White Dwarf-main Sequence Binaries}",
      journal = {\pasp},
         year = 2026,
        month = mar,
       volume = {138},
       number = {3},
          eid = {034202},
        pages = {034202},
          doi = {10.1088/1538-3873/ae453b},
archivePrefix = {arXiv},
       eprint = {2601.00439},
 primaryClass = {astro-ph.SR},
       adsurl = {https://ui.adsabs.harvard.edu/abs/2026PASP..138c4202S}
}

@ARTICLE{1981A&A....99..126H,
       author = {{Hut}, P.},
        title = "{Tidal evolution in close binary systems.}",
      journal = {\aap},
         year = 1981,
        month = jun,
       volume = {99},
        pages = {126-140},
       adsurl = {https://ui.adsabs.harvard.edu/abs/1981A&A....99..126H}
}

@ARTICLE{2014A&A...566A..86C,
       author = {{Camacho}, Judit and {Torres}, Santiago and {Garc{\'\i}a-Berro}, Enrique and {Zorotovic}, M{\'o}nica and {Schreiber}, Matthias R. and {Rebassa-Mansergas}, Alberto and {Nebot G{\'o}mez-Mor{\'a}n}, Ada and {G{\"a}nsicke}, Boris T.},
        title = "{Monte Carlo simulations of post-common-envelope white dwarf + main sequence binaries: comparison with the SDSS DR7 observed sample}",
      journal = {\aap},
         year = 2014,
        month = jun,
       volume = {566},
          eid = {A86},
        pages = {A86},
          doi = {10.1051/0004-6361/201323052},
archivePrefix = {arXiv},
       eprint = {1404.5464},
 primaryClass = {astro-ph.GA},
       adsurl = {https://ui.adsabs.harvard.edu/abs/2014A&A...566A..86C}
}

@ARTICLE{2017MNRAS.470.1442C,
       author = {{Cojocaru}, R. and {Rebassa-Mansergas}, A. and {Torres}, S. and {Garc{\'\i}a-Berro}, E.},
        title = "{The population of white dwarf-main sequence binaries in the SDSS DR 12}",
      journal = {\mnras},
         year = 2017,
        month = sep,
       volume = {470},
       number = {2},
        pages = {1442-1452},
          doi = {10.1093/mnras/stx1326},
archivePrefix = {arXiv},
       eprint = {1705.05888},
 primaryClass = {astro-ph.SR},
       adsurl = {https://ui.adsabs.harvard.edu/abs/2017MNRAS.470.1442C}
}

@ARTICLE{2025A&A...698A.173T,
       author = {{Torres}, S. and {Gili}, M. and {Rebassa-Mansergas}, A. and {Santos-Garc{\'\i}a}, A. and {Brown}, A.~J. and {Parsons}, S.~G.},
        title = "{Reconstructing post-common envelope white dwarf─main-sequence binary histories through inverse population synthesis techniques: A case study of ZTF eclipsing binaries}",
      journal = {\aap},
         year = 2025,
        month = jun,
       volume = {698},
          eid = {A173},
        pages = {A173},
          doi = {10.1051/0004-6361/202554039},
archivePrefix = {arXiv},
       eprint = {2505.01505},
 primaryClass = {astro-ph.SR},
       adsurl = {https://ui.adsabs.harvard.edu/abs/2025A&A...698A.173T}
}

@ARTICLE{2016MNRAS.458.3808R,
       author = {{Rebassa-Mansergas}, A. and {Ren}, J.~J. and {Parsons}, S.~G. and {G{\"a}nsicke}, B.~T. and {Schreiber}, M.~R. and {Garc{\'\i}a-Berro}, E. and {Liu}, X.-W. and {Koester}, D.},
        title = "{The SDSS spectroscopic catalogue of white dwarf-main-sequence binaries: new identifications from DR 9-12}",
      journal = {\mnras},
         year = 2016,
        month = jun,
       volume = {458},
       number = {4},
        pages = {3808-3819},
          doi = {10.1093/mnras/stw554},
archivePrefix = {arXiv},
       eprint = {1603.01017},
 primaryClass = {astro-ph.SR},
       adsurl = {https://ui.adsabs.harvard.edu/abs/2016MNRAS.458.3808R}
}

@ARTICLE{2016MNRAS.463.2125P,
       author = {{Parsons}, S.~G. and {Rebassa-Mansergas}, A. and {Schreiber}, M.~R. and {G{\"a}nsicke}, B.~T. and {Zorotovic}, M. and {Ren}, J.~J.},
        title = "{The white dwarf binary pathways survey - I. A sample of FGK stars with white dwarf companions}",
      journal = {\mnras},
         year = 2016,
        month = dec,
       volume = {463},
       number = {2},
        pages = {2125-2136},
          doi = {10.1093/mnras/stw2143},
archivePrefix = {arXiv},
       eprint = {1604.01613},
 primaryClass = {astro-ph.SR},
       adsurl = {https://ui.adsabs.harvard.edu/abs/2016MNRAS.463.2125P}
}

@ARTICLE{2017MNRAS.472.4193R,
       author = {{Rebassa-Mansergas}, A. and {Ren}, J.~J. and {Irawati}, P. and {Garc{\'\i}a-Berro}, E. and {Parsons}, S.~G. and {Schreiber}, M.~R. and {G{\"a}nsicke}, B.~T. and {Rodr{\'\i}guez-Gil}, P. and {Liu}, X. and {Manser}, C. and {Nevado}, S.~P. and {Jim{\'e}nez-Ibarra}, F. and {Costero}, R. and {Echevarr{\'\i}a}, J. and {Michel}, R. and {Zorotovic}, M. and {Hollands}, M. and {Han}, Z. and {Luo}, A. and {Villaver}, E. and {Kong}, X.},
        title = "{The white dwarf binary pathways survey - II. Radial velocities of 1453 FGK stars with white dwarf companions from LAMOST DR 4}",
      journal = {\mnras},
         year = 2017,
        month = dec,
       volume = {472},
       number = {4},
        pages = {4193-4203},
          doi = {10.1093/mnras/stx2259},
archivePrefix = {arXiv},
       eprint = {1708.09480},
 primaryClass = {astro-ph.SR},
       adsurl = {https://ui.adsabs.harvard.edu/abs/2017MNRAS.472.4193R}
}

@ARTICLE{2020ApJ...905...38R,
       author = {{Ren}, J.-J. and {Raddi}, R. and {Rebassa-Mansergas}, A. and {Hernandez}, M.~S. and {Parsons}, S.~G. and {Irawati}, P. and {Rittipruk}, P. and {Schreiber}, M.~R. and {G{\"a}nsicke}, B.~T. and {Torres}, S. and {Wang}, H.-J. and {Zhang}, J.-B. and {Zhao}, Y. and {Zhou}, Y.-T. and {Han}, Z.-W. and {Wang}, B. and {Liu}, C. and {Liu}, X.-W. and {Wang}, Y. and {Zheng}, J. and {Wang}, J.-F. and {Zhao}, F. and {Cui}, K.-M. and {Shi}, J.-R. and {Tian}, H.},
        title = "{The White Dwarf Binary Pathways Survey. V. The Gaia White Dwarf Plus AFGK Binary Sample and the Identification of 23 Close Binaries}",
      journal = {\apj},
         year = 2020,
        month = dec,
       volume = {905},
       number = {1},
          eid = {38},
        pages = {38},
          doi = {10.3847/1538-4357/abc017},
archivePrefix = {arXiv},
       eprint = {2010.02885},
 primaryClass = {astro-ph.SR},
       adsurl = {https://ui.adsabs.harvard.edu/abs/2020ApJ...905...38R}
}

@ARTICLE{2026ApJ..1003..145B,
       author = {{Boone}, Alexandra and {Kobulnicky}, Henry A. and {Ca{\~n}as}, Caleb I. and {Kanodia}, Shubham and {Monson}, Andrew and {Shea}, Peter and {Cochran}, William and {Mahadevan}, Suvrath and {Ninan}, Joe and {Robertson}, Paul and {Han}, Te and {Roy}, Arpita and {Schwab}, Christian and {Allen}, Madeleine},
        title = "{Searching for GEMS: Discovery of the Nearby Post-common-envelope Binary System TIC-460388167}",
      journal = {\apj},
         year = 2026,
        month = jun,
       volume = {1003},
       number = {2},
          eid = {145},
        pages = {145},
          doi = {10.3847/1538-4357/ae6246},
archivePrefix = {arXiv},
       eprint = {2604.07527},
 primaryClass = {astro-ph.SR},
       adsurl = {https://ui.adsabs.harvard.edu/abs/2026ApJ..1003..145B}
}

@ARTICLE{2022MNRAS.516.1183P,
       author = {{Priyatikanto}, R. and {Knigge}, C. and {Scaringi}, S. and {Brink}, J. and {Buckley}, D.~A.~H.},
        title = "{Characterization of the eclipsing post-common-envelope binary TIC 60040774}",
      journal = {\mnras},
         year = 2022,
        month = oct,
       volume = {516},
       number = {1},
        pages = {1183-1192},
          doi = {10.1093/mnras/stac2197},
archivePrefix = {arXiv},
       eprint = {2208.02986},
 primaryClass = {astro-ph.SR},
       adsurl = {https://ui.adsabs.harvard.edu/abs/2022MNRAS.516.1183P}
}

@ARTICLE{1994MNRAS.270..121H,
       author = {{Han}, Z. and {Podsiadlowski}, P. and {Eggleton}, P.~P.},
        title = "{A possible criterion for envelope ejection in asymptotic giant branch or first giant branch stars.}",
      journal = {\mnras},
         year = 1994,
        month = sep,
       volume = {270},
        pages = {121-130},
          doi = {10.1093/mnras/270.1.121},
       adsurl = {https://ui.adsabs.harvard.edu/abs/1994MNRAS.270..121H}
}

@ARTICLE{2014Sci...344..275K,
       author = {{Kruse}, Ethan and {Agol}, Eric},
        title = "{KOI-3278: A Self-Lensing Binary Star System}",
      journal = {Science},
         year = 2014,
        month = apr,
       volume = {344},
       number = {6181},
        pages = {275-277},
          doi = {10.1126/science.1251999},
archivePrefix = {arXiv},
       eprint = {1404.4379},
 primaryClass = {astro-ph.SR},
       adsurl = {https://ui.adsabs.harvard.edu/abs/2014Sci...344..275K}
}

@ARTICLE{2019MNRAS.484.5362A,
       author = {{Ashley}, R.~P. and {Farihi}, J. and {Marsh}, T.~R. and {Wilson}, D.~J. and {G{\"a}nsicke}, B.~T.},
        title = "{Evidence for bimodal orbital separations of white dwarf-red dwarf binary stars}",
      journal = {\mnras},
         year = 2019,
        month = apr,
       volume = {484},
       number = {4},
        pages = {5362-5376},
          doi = {10.1093/mnras/stz298},
archivePrefix = {arXiv},
       eprint = {1901.09139},
 primaryClass = {astro-ph.SR},
       adsurl = {https://ui.adsabs.harvard.edu/abs/2019MNRAS.484.5362A}
}

@ARTICLE{2022MNRAS.512.2625L,
       author = {{Lagos}, F. and {Schreiber}, M.~R. and {Parsons}, S.~G. and {Toloza}, O. and {G{\"a}nsicke}, B.~T. and {Hernandez}, M.~S. and {Schmidtobreick}, L. and {Belloni}, D.},
        title = "{The white dwarf binary pathways survey - VII. Evidence for a bi-modal distribution of post-mass transfer systems?}",
      journal = {\mnras},
         year = 2022,
        month = may,
       volume = {512},
       number = {2},
        pages = {2625-2635},
          doi = {10.1093/mnras/stac673},
archivePrefix = {arXiv},
       eprint = {2203.08846},
 primaryClass = {astro-ph.SR},
       adsurl = {https://ui.adsabs.harvard.edu/abs/2022MNRAS.512.2625L}
}

@misc{zheng2026massorbitalperioddistributionmassive,
      title={Mass-Orbital Period Distribution of Massive White Dwarfs Formed Through Stable Mass Transfer}, 
      author={Rizhong Zheng and Hongwei Ge and Christopher A Tout and Hailiang Chen and Zhenwei Li and Dengkai Jiang and Chengyuan Li and Zhijia Tian and Bo Ma and Lifu Zhang and Jian Mou and Xuefei Chen and Zhanwen Han},
      year={2026},
      eprint={2606.06141},
      archivePrefix={arXiv},
      primaryClass={astro-ph.SR},
      url={https://arxiv.org/abs/2606.06141}, 
}

@ARTICLE{2013A&A...559A.104T,
       author = {{Tremblay}, P.-E. and {Ludwig}, H.-G. and {Steffen}, M. and {Freytag}, B.},
        title = "{Spectroscopic analysis of DA white dwarfs with 3D model atmospheres}",
      journal = {\aap},
         year = 2013,
        month = nov,
       volume = {559},
          eid = {A104},
        pages = {A104},
          doi = {10.1051/0004-6361/201322318},
archivePrefix = {arXiv},
       eprint = {1309.0886},
 primaryClass = {astro-ph.SR},
       adsurl = {https://ui.adsabs.harvard.edu/abs/2013A&A...559A.104T}
}

@ARTICLE{2015ApJ...800..114S,
       author = {{Soker}, Noam},
        title = "{Close Stellar Binary Systems by Grazing Envelope Evolution}",
      journal = {\apj},
         year = 2015,
        month = feb,
       volume = {800},
       number = {2},
          eid = {114},
        pages = {114},
          doi = {10.1088/0004-637X/800/2/114},
archivePrefix = {arXiv},
       eprint = {1410.5363},
 primaryClass = {astro-ph.SR},
       adsurl = {https://ui.adsabs.harvard.edu/abs/2015ApJ...800..114S}
}

@ARTICLE{2018MNRAS.480.3195K,
       author = {{Kashi}, Amit and {Soker}, Noam},
        title = "{Counteracting tidal circularization with the grazing envelope evolution}",
      journal = {\mnras},
         year = 2018,
        month = nov,
       volume = {480},
       number = {3},
        pages = {3195-3200},
          doi = {10.1093/mnras/sty2115},
archivePrefix = {arXiv},
       eprint = {1807.03042},
 primaryClass = {astro-ph.SR},
       adsurl = {https://ui.adsabs.harvard.edu/abs/2018MNRAS.480.3195K}
}

@ARTICLE{2018MNRAS.477.2584S,
       author = {{Shiber}, Sagiv and {Soker}, Noam},
        title = "{Simulating a binary system that experiences the grazing envelope evolution}",
      journal = {\mnras},
         year = 2018,
        month = jun,
       volume = {477},
       number = {2},
        pages = {2584-2598},
          doi = {10.1093/mnras/sty843},
archivePrefix = {arXiv},
       eprint = {1706.00398},
 primaryClass = {astro-ph.SR},
       adsurl = {https://ui.adsabs.harvard.edu/abs/2018MNRAS.477.2584S}
}

@ARTICLE{2025A&A...698A.257I,
       author = {{I{\l}kiewicz}, Krystian and {Planquart}, L{\'e}a and {Kami{\'n}ski}, Tomek and {Van Winckel}, Hans},
        title = "{Circumbinary disc interactions and stochastic dust obscuration in the post-asymptotic-giant-branch binary HD 213985}",
      journal = {\aap},
         year = 2025,
        month = jun,
       volume = {698},
          eid = {A257},
        pages = {A257},
          doi = {10.1051/0004-6361/202453585},
archivePrefix = {arXiv},
       eprint = {2506.07260},
 primaryClass = {astro-ph.SR},
       adsurl = {https://ui.adsabs.harvard.edu/abs/2025A&A...698A.257I}
}

@ARTICLE{2026A&A...707A.342C,
       author = {{Cui}, Yingzhen and {Wang}, Song and {Meng}, Xiangcun and {Liu}, Jifeng and {Ma}, Shuguo and {Zhao}, Weitao},
        title = "{Dynamical mass loss at the end of thermally pulsating asymptotic giant branch stars}",
      journal = {\aap},
         year = 2026,
        month = mar,
       volume = {707},
          eid = {A342},
        pages = {A342},
          doi = {10.1051/0004-6361/202556962},
archivePrefix = {arXiv},
       eprint = {2601.18194},
 primaryClass = {astro-ph.SR},
       adsurl = {https://ui.adsabs.harvard.edu/abs/2026A&A...707A.342C}
}
\bibliographystyle{aasjournalv7}



\end{document}